\documentclass[a4paper,fleqn]{cas-dc}

\usepackage[authoryear]{natbib}

\usepackage{amsmath,amsfonts}
\usepackage{array}
\usepackage{textcomp}
\usepackage{graphicx}
\usepackage{url}
\usepackage{colortbl}
\definecolor{bestgreen}{HTML}{B9DCA4}
\definecolor{neargreen}{HTML}{E6F2DE}
\definecolor{lowgray}{gray}{0.55}

\DeclareUnicodeCharacter{2011}{-}
\DeclareUnicodeCharacter{00D7}{\ensuremath{\times}}

\begin{document}
\let\WriteBookmarks\relax
\def\floatpagepagefraction{1}
\def\textpagefraction{.001}
\renewcommand{\topfraction}{0.95}
\renewcommand{\dbltopfraction}{0.95}
\renewcommand{\floatpagefraction}{0.5}
\renewcommand{\dblfloatpagefraction}{0.5}
\shorttitle{Dual-timescale memory in a spiking neuron--astrocyte network}
\shortauthors{Yu. Tsybina et~al.}

\title [mode = title]{Neuromorphic Navigation in Static and Dynamic Environments with Dual-Timescale Neuron-Astrocyte Memory}

\author[1,2]{Yuliya Tsybina}[orcid=0000-0002-8463-1226]
\ead{lotarevaj@gmail.com}
\credit{Methodology, Software, Formal analysis, Investigation, Visualization, Writing -- original draft}

\author[1]{Evgenia Antonova}[orcid=0009-0000-2700-6902]
\ead{jenya.antonova@bk.ru}
\credit{Investigation, Data curation, Formal analysis}

\author[1]{Sergey Shchanikov}[orcid=0000-0002-3938-8896]
\ead{seach@inbox.ru}
\credit{Software, Investigation (hardware implementation), Formal analysis, Writing -- original draft}

\author[3]{Vsevolod Kulagin}[orcid=0000-0003-3397-7910]
\ead{kulagin.V.A@outlook.com}
\credit{Methodology, Software, Formal analysis}

\author[1]{Alexey Mikhaylov}[orcid=0000-0001-5505-7352]
\ead{mian@nifti.unn.ru}
\credit{Resources, Writing -- review \& editing}

\author[1,2]{Victor Kazantsev}[orcid=0000-0002-2881-6648]
\ead{kazantsev@neuro.nnov.ru}
\credit{Writing -- review \& editing}

\author[3]{Vyacheslav Demin}[orcid=0000-0001-9142-4295]
\ead{demin_va@nrcki.ru}
\credit{Conceptualization, Methodology, Supervision, Writing -- original draft, Writing -- review \& editing}

\author[1,2]{Susanna Gordleeva}[orcid=0000-0002-7687-3065]
\cormark[1]
\ead{gordleeva@neuro.nnov.ru}
\credit{Conceptualization, Methodology, Supervision, Writing -- original draft, Writing -- review \& editing}

\affiliation[1]{organization={Lobachevsky State University of Nizhny Novgorod},
                city={Nizhny Novgorod},
                country={Russia}}

\affiliation[2]{organization={Neuromorphic Computing Center, Neimark University},
                city={Nizhny Novgorod},
                country={Russia}}

\affiliation[3]{organization={National Research Center ``Kurchatov Institute''},
                city={Moscow},
                country={Russia}}

\cortext[1]{Corresponding author}

\begin{abstract}
Biological agents navigate complex environments by combining long-term memory of successful actions with short-term suppression of recently visited locations---a capability difficult to replicate in artificial systems. We introduce a neuromorphic navigation system based on a spiking neuron--astrocyte network (SNAN), in which spike-timing-dependent plasticity (STDP) reinforces traversed action sequences on a long timescale, while astrocytic calcium transients suppress moves into recently visited states on a short timescale. This dual-timescale memory biases the agent toward unexplored regions without reward signals or explicit global statistics. In grid worlds where the agent observes only its current position, astrocytic modulation increased goal completion from 45.9\% to 99.9\% and shortened median routes 5.8–15-fold on commonly solved tasks. Compared with seven classical strategies, including Tr\'emaux's and Tarry's algorithms and tabular Q-learning, SNAN yields the shortest routes among reliable strategies in open and sparse static mazes while storing 2.8–4.6 times fewer variables than memory-based baselines. In dynamic mazes, its decaying memory retains no obsolete passage marks, so SNAN becomes more efficient as the topology changes. After ten learning cycles, STDP makes SNAN the best reliable strategy. The astrocytic modulation mitigates the exploration-exploitation trade-off through local state suppression, a form of working memory we term Topological-Context Memory. To validate hardware feasibility, we map STDP to a memristive VTEAM model and implement a subset of the network on a crossbar array, achieving order-of-magnitude gains in speed per area and energy per decision over CPU implementations. These results establish astrocyte-inspired dual-timescale memory as a compact, hardware-compatible principle for neuromorphic navigation.

\end{abstract}

% Research highlights (3-5 bullets, max 85 characters each incl. spaces).
\begin{highlights}
\item Astrocytes underlie short-term memory of positions visited by neuromorphic agents.
\item Two timescales, one network: STDP exploits, astrocytic calcium drives exploration.
\item With astrocytes, goal completion jumps from 45.9\% to 99.9\% and routes shrink 5.8–15×.
\item Forgetting helps: astrocytic memory outpaces Tr\'emaux and Tarry in changing mazes.
\item Memristive crossbar cuts energy per decision by orders of magnitude versus a CPU.
\end{highlights}


\maketitle

\section{Introduction}\label{sec_Intro}
The problem of balancing exploration and exploitation in an agent's optimal behavior strategy has a long history in operant conditioning \citep{Akpan2020, Staddon2003} and machine learning \citep{Barto1981}. On the one hand, without exploration, i.e., iterative changes to its strategy, the agent may be far from the optimal strategy. The optimal strategy is defined through demonstrating a behavior that accumulates the maximum cumulative value of some given reward or penalty function in the certain environment. Moreover, insufficiently active exploration of different behavior patterns will not allow the agent to converge to an optimal strategy, since a necessary condition for such convergence is infinite visitation of each state of the environment \citep{Singh2000}. On the other hand, however, this condition represents an extreme degree of bias towards exploration in the exploration-exploitation dilemma and is practically unattainable in almost all practical cases.

Therefore, in the absence of using suboptimal but sometimes acceptable behavioral strategies, e.g., when actions are chosen randomly by the agent, solving the task can take significant time or even make it impossible altogether (for example, if the agent gets stuck in a limited state space due to looping, algorithm termination, or ``death''). Indeed, good but non-optimal strategies enable collecting high-quality data about the structure of the optimized function within the state-action space available to the agent, which can then be used to improve the strategy at subsequent iterations. This exploitation tactic allows iteratively finding near-optimal solutions even in very complex and weakly structured environments, including physical artificial intelligence \citep{Schneider2024}, autonomous agents and virtual environments \citep{Tihomirov2025, Oh2015, Vinyals2019, Vlasov2023}, linguistic spaces for large language models \citep{lee2023}, etc.

But how to determine the optimal balance between exploring new behaviors and exploiting already known good strategies for intelligent agents? Theoretically, this problem has been solved only in a very limited number of tasks \citep[e.g.,][]{Slivkins2019, Auer2002}, while in most practical cases it is addressed heuristically. In many algorithms from reinforcement learning and large generative language model domains, this issue is determined via hyperparameters of the algorithm, such as mini-batch size of collected data, numbers of parallel running environments/experts for data collection, learning rate coefficient, etc. \citep{Murphy2025}.

Even more challenging and understudied appears the considered problem in the context of brain functioning in animals and humans: what biological mechanisms underlie the establishment and modulation of the search-exploit balance for behavioral strategies of living organisms? The challenge is exacerbated by the fact that neural network activation and information transmission mechanisms typically operate on timescales ranging from single units to tens of milliseconds, whereas organismal behavior is determined over scales from hundreds of milliseconds to seconds and beyond (at least for most mammals). How do neuron activation patterns link into chains of actions or reasoning lasting several seconds or longer? If action distribution learning from current states $\pi$($a|s$) is employed, then either the external environmental states must carry exhaustive information for decision-making (approximation of Markov Decision Processes \citep{Auer2002}) or there needs to be a mechanism for retaining recent past information for use in present-time decisions.

Indeed, a classical approach to addressing the problem of establishing a balance between searching and exploiting a strategy involves incorporating memory mechanisms into the model, allowing consideration of recent historical contexts. This is particularly important in partially observable environments where a complete description of the current state is unavailable to the agent. In such scenarios, the agent should remember what it ``saw'' one step back, two steps back, and so forth because these observations critically influence the choice of the optimal action at the current step. Thus, the agent’s strategy transforms into the form $\pi$($a_t|o_t, o_{t-1},\ldots, o_0$), where $o_t$ denotes a partial observation of the state (such as observing just one cell in a grid-like environment or maze), and generally speaking, the agent has access to all observations starting from the first episode $o_0$. Sometimes, the previous context is encoded by means of a latent vector $h_{t-1}$ = $\varphi$($o_{t-1},\ldots, o_0$), and the policy $\pi(a_t|o_t, h_{t-1})$ is trained, for instance, using recurrent neural networks with input ($o_t$, $h_{t-1}$), outputting both $h_t$ and $a_t$ at every iteration \citep{Greff2017}. As is well known, the hippocampus—a paired structure of the limbic system in higher animal brains and human brains—plays a key role in forming long-term memories \citep{Eichenbaum2000}. At the same time, the mechanisms of short-term (working) memory operating at the relevant temporal level, from fractions to dozens of seconds, have been studied less extensively. Both dynamic principles of working memory operation, manifested as stable attractors and repellers of neuronal activity spike patterns \citep{Kazantsev2005}, and, recently, mechanisms involving short-term synaptic plasticity mediated by glial cells (astrocytes) supporting computations by neurons across appropriate time ranges through slow calcium waves generated by astrocytes, have been proposed. It was demonstrated that integrating experimentally validated mechanisms of astrocyte-mediated synaptic transmission modulation into biologically relevant spiking neural network models enables formation and reliable maintenance of short-term memory on the timescale of astrocytic calcium activity \citep{Gordleeva2021, Gordleeva2025}.

In this work, we investigate the possibility of engaging astrocytes in providing operational memory mechanisms to address the problem of maintaining a balance between exploration and exploitation of an agent's strategy using a model grid-based environment where only one cell is observed at any moment in time. We further compare the proposed neuromorphic navigation system with classical navigation strategies---random walk, random mouse, wall follower, Pledge, Tr\'emaux's and Tarry's algorithms, and tabular Q-learning---in both static mazes and dynamic mazes whose topology changes during navigation. This comparison evaluates not only route length and success rate but also memory requirements and the ability to reuse experience. Furthermore, this problem is also explored from the perspective of hardware implementation of neuromorphic learning mechanisms based on memristors, nonvolatile resistive memory cells acting as analogues of synaptic connections between artificial neurons \citep{Chua1971, Strukov2008, Kulagin2025, Mikhaylov2020, Rybka2024, Vlasov2024}. Such a formulation of the problem raises great interest, as it not only enhances processor device characteristics for machine learning applications regarding energy efficiency and low computational latency \citep{Benjamin2014}, but also opens up possibilities for continuous in situ online training of an agent in real-time. This potentially broadens prospects for robotics, wearable smart assistants, devices for body-worn medicine, and other smart gadgets adapting to user habits and needs during their usage \citep{Thrun1995}.

The rest of the paper is organized as follows: Section~\ref{sec_Methods} provides detailed information concerning the task, biologically inspired agent models, and their hardware implementation on memristive electronics. The experimental setup and results of the experiments performed on 2D grid world environments are described in Section~\ref{sec_results}, and the hardware implementation in Section~\ref{sec_hardware}. Finally, Sections~\ref{sec_discussion} and~\ref{sec_conclusion} discuss the results, outline the limitations of the proposed method, and provide insights for future work.

\section{Methods}\label{sec_Methods}

\subsection{Problem Statement}\label{problem}
\begin{figure*}[t]
\centering
\includegraphics[width=\textwidth]{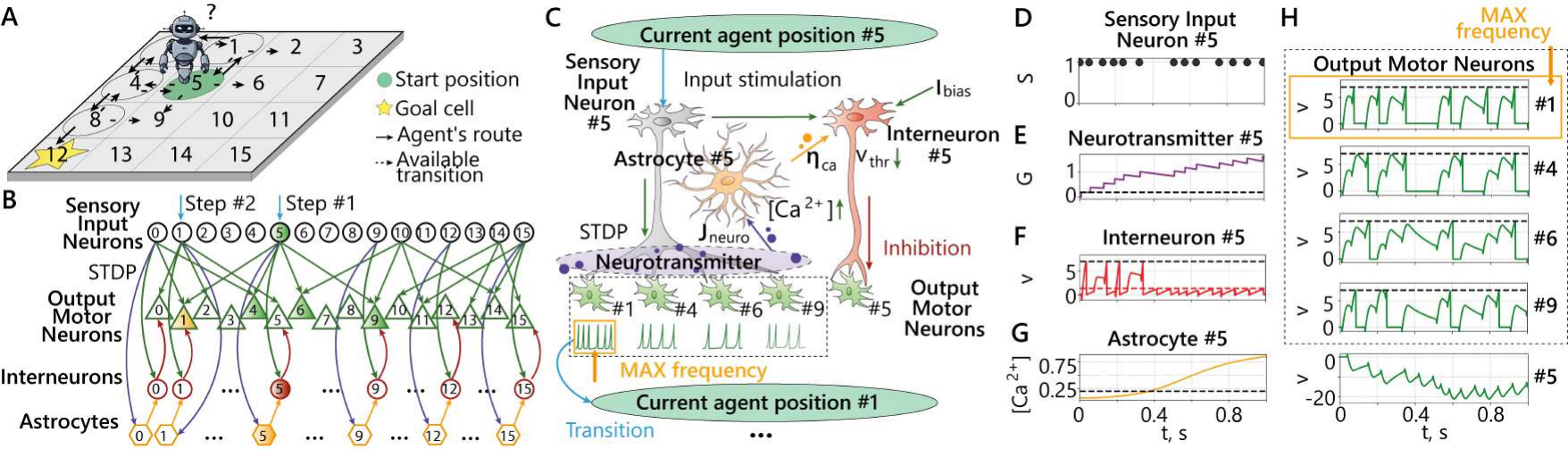}
\caption{The considered grid environment and the proposed bioinspired short-term memory model implemented by a spiking neuron–astrocyte network (SNAN) for an agent navigating.
(A) An agent navigating a discrete 2D grid from an initial position (green cell) to a goal (yellow cell). Arrows indicate possible movement paths along the cardinal directions. (B) SNAN architecture. (C) Schematic of the SNAN at the cellular level. (D–H) Dynamics of key variables: (D) Spike raster of sensory input neuron \#5, (E) Neurotransmitter released into the extracellular space by sensory input neuron \#5, (F) Membrane potential of interneuron \#5, (G) Intracellular calcium concentration in the interacting astrocyte \#5, and (H) Membrane potential of output motor neurons \#1,4,6,9 synaptically connected to sensory input neuron \#5.}
\label{fig1}
\end{figure*}

We consider the problem of navigating an agent in a partially observable grid environment, where the agent lacks complete, real-time access to the full state of its surroundings, except for an observation that consists of a single cell of a grid environment denoting its position (Fig.~\ref{fig1}A). Each state can be defined by the agent’s position; thus, state space $S$ contains all possible agent positions. The environment’s transition function is deterministic. The action space consists of four actions $A$ that move the agent to each adjacent grid cell: up, down, left, and right. The agent receives no observations about the environment. However, the agent is unable to move beyond the boundaries of the environment or through the maze walls. At the beginning of each test trial, the agent's initial state, $S_{init}$, and goal state, $S_{G}$ are chosen randomly, with $S_{init} \neq S_G$. The trial ends when the agent reaches the goal state or when the time limit is reached. As a metric of the agent’s performance during a trial, we use the number of steps required for the agent to reach the goal. For each agent and environment configuration, we perform several independent trials.

\subsection{Spiking Neuron-Astrocyte Network Model}\label{network1}

We propose a bioinspired short-term memory model implemented by a spiking neuron–astrocyte network (SNAN) of an agent operating in an unfamiliar grid environment. The concept of the proposed SNAN of an intelligent agent is schematically summarized in Fig.~\ref{fig1}. The proposed SNAN is a neurophysiologically inspired hybrid system for balancing exploration and exploitation strategies in agents. It combines a fast-spiking neural network, which is responsible for taking actions to achieve goals (exploitation strategy), and a slow astrocytic network. This astrocytic network provides time-dependent data buffering via calcium activity and gliotransmitter-induced spatiotemporal coordination of neural network activity, which is responsible for remembering several past observations to reduce uncertainty about the environment (exploration strategy). Indeed, converging experimental evidence from diverse CNS circuits and behavioral paradigms demonstrates that astrocytes are actively and causally involved in information processing and memory formation across multiple spatiotemporal scales, from local synaptic plasticity and network oscillations to cognitive performance under both physiological and pathological conditions \citep{Santello2019}.

\subsubsection{SNAN Architecture}\label{network1_0}
The architecture of the proposed simple SNAN is shown in Fig.~\ref{fig1}B-C. The SNAN consists of interacting units, the number of which is determined by the number of agent states or the number of cells in the grid environment. Schematically, each unit of SNAN consists of one sensory input neuron, one output motor (pyramidal) neuron, one interneuron, and one astrocyte. The synaptic interactions between cells within a single unit occur as follows: each cell of the grid environment corresponds to its input neuron. When the agent occupies a cell on the grid, its corresponding input neuron generates spikes. The input neuron stimulates the motor neurons corresponding to neighboring cells on the grid. The synaptic connections between the input neuron and the motor neurons are plastic and are modified according to the STDP rule. All other synaptic connections within each unit are non-plastic. The next position of the agent is determined by the motor neuron that exhibits the highest spike frequency. The activity of the input neuron activates also the inhibitory neuron and the astrocyte within that unit. The inhibitory neuron is synaptically connected to the motor neuron, thereby suppressing its activity. It is important to note that each inhibitory interneuron is also tonically stimulated by some fixed constant current coming from some pool of pacemaker neurons (not shown) with fixed but astrocyte-modulated connections to the interneuron. The astrocyte tracks the activity of input neuron and responds to it with slow intracellular calcium elevations, which trigger the release of gliotransmitters. Gliotransmitter-induced short-term synaptic plasticity results in enhanced excitability of inhibitory neuron  and consequently enhances its inhibitory action on the motor neuron in that unit.  Thus, in the proposed agent navigation system, short-term memory of the agent's current and several past states (grid positions) is implemented via astrocytic regulation of neural activity. The Ca-activated astrocyte, through acting on synaptic connections with interneuron and pacemaker neurons (not shown), suppresses the activity of an output neuron for a period of time, thereby preventing the agent from selecting the grid cell corresponding to that neuron as its next position. This modulation operates on a timescale of seconds, which is determined by the biophysical mechanisms underlying the generation and duration of the calcium transient in the astrocyte and the consequent astrocytic modulation of synaptic transmission. Detailed information concerning the models and the description of parameters is provided in Sections~\ref{network1_1}-\ref{network1_4}.

\subsubsection{Neural Network Model}\label{network1_1}

Presented schematics of SNAN can be further specialized as follows. When the agent is located in cell $i$ of the grid environment, the corresponding $i$-th input neuron generates a spike train $S_{i}$ over a period $T_{stim}$ according to the following rule: within each timestep, a spike is emitted with a fixed probability $p$, implementing a Bernoulli process \citep{Edwards1990} (Fig.~\ref{fig1}D). The input neurons associated with unoccupied grid cells remain silent.

These $i$-th input neuron activates both the $i$-th interneuron and the motor neurons. The input-to-output motor connections follow a local adjacency pattern, where each input neuron selectively targets only output motor neurons representing neighboring grid cells. For example, if the agent's current position is cell 5 (see Fig.~\ref{fig1}A), then the 5-th input neuron activates the motor neurons numbered 1, 4, 6, and 9.

The motor neurons (Fig.~\ref{fig1}H) implement discrete leaky integrate-and-fire (LIF) dynamics \citep{Zenke2018} described by:

\begin{equation}
\label{eq:LIF_excitatory}
\begin{aligned}
v_{j}[t] &= v_{\text{rest}} + \alpha (v_{j}[t-\Delta t] - v_{\text{rest}}) + R \cdot I_{\text{syn}j}[t], \\
\alpha &= \exp(-\Delta t / \tau_m), \\
I_{\text{syn}j}[t] &= w_{ij} \cdot S_{i}[t] - w_{I} \cdot S_{Ij}[t],\\
&\textbf{if } v_{j}[t] > v_{thr} \textbf{ then } v_{j}[t] = v_{\text{reset}},
\end{aligned}
\end{equation}

\noindent where $j$ denotes the motor neuron index, t is time in milliseconds, $v_{j}[t]$ is the membrane potential of motor neuron $j$ at time $t$, $v_{\text{rest}}$ is the resting potential, $\alpha$ is the membrane potential decay factor, $\Delta t$ is the simulation timestep (1 ms), $\tau_m$ is the membrane time constant, $R$ is the membrane resistance, and $I_{\text{syn}j}[t]$ is the input synaptic current. $w_{ij}$ is the matrix of plastic synaptic connection weights from input neurons to motor neurons, and $w_{I}$ is the fixed identity matrix of synaptic connection weights from interneurons to motor neurons. $S_{i}[t]$ and $S_{Ij}[t]$ denote activity of the $i$-th input neuron and the $j$-th interneuron, respectively, at time $t$. $v_{thr}$ and $v_{reset}$ are the firing threshold and the post-spike reset potential, respectively.

Interneurons (Fig.~\ref{fig1}F) follow the same LIF dynamics, but with an additional constant external bias current $I_{\text{bias}}$ added to the right-hand side of the membrane potential equation. The input synaptic current is provided by the input neurons through fixed one-to-one connections, so that each interneuron receives spikes only from its corresponding input neuron.

All neuron types implement an absolute refractory period. Following a spike, each neuron enters a refractory state for a fixed duration $t_{ref}$. During this period, the neuron cannot receive signals or generate spikes. A detailed list of parameter values can be found in Table~\ref{tab_1}.

\textit{Spike-timing-dependent plasticity (STDP)}

In the proposed SNAN, the synaptic weights from input to output motor neurons, $w_{ij}$, dynamically adjust during agent navigation in grid environment. The initial weights of the synapses between input and motor neurons are set from a normal distribution with mean $\mu_w$ and standard deviation $\sigma_w$. Updated weights are constrained to the fixed range [$w_{min}$, $w_{max}$]. The additive STDP rule updates the synaptic weights according to the timing difference between the pre- and postsynaptic spikes ($\Delta T$ = $t_{post}$ - $t_{pre}$) and is described by:

\begin{equation}
\label{eq:stdp}
\Delta w_{ij} =
\begin{cases}
A_+ \cdot \exp\left(-\frac{\Delta T}{\tau_+}\right), & \text{if } \Delta T \geq 0; \\
-A_- \cdot \exp\left(-\frac{|\Delta T|}{\tau_-}\right), & \text{if } \Delta T < 0;
\end{cases}
\end{equation}

\noindent where $\Delta w_{ij}$ is used to update the synaptic weight, $A_+$ and $A_-$ control the amplitudes of potentiation and depression, while $\tau_+$ and $\tau_-$ are their respective time constants. To reinforce the pathway corresponding to the agent's actual move, the cumulative STDP weight change for the synapse from the current input neuron to the motor neuron of the chosen next grid cell was additionally multiplied by a factor $k$.

The agent selects its next movement direction based on the activity of motor neurons corresponding to adjacent grid cells from its current position. At the end of each stimulation period $T_{stim}$, the agent moves to the grid cell corresponding to the motor neuron with the highest spiking frequency during this decision interval (Fig.~\ref{fig1}C,H). If multiple neurons have the same maximum spike count, the next position is chosen randomly from among these candidates. The agent's new position is then encoded by the $(i+1)$-th input neuron, which generates its spike train $S_{i+1}$ over the next stimulation period $T_{\text{stim}}$ as described above. This loop repeats iteratively until the agent reaches the goal.

\subsubsection{Astrocyte Model}\label{network1_3}

Spiking input neuronal activity induces the release of neurotransmitters (glutamate) into the extracellular space (Fig.~\ref{fig1}E). The amount of neurotransmitter glutamate that diffuses from the synaptic cleft associated with the $i$-th input neuron and reaches the astrocyte is described by the following equation \citep{Gordleeva2012}:

\begin{equation}
\frac{dG_i}{dt} = -\alpha_{\text{n}} G_i + k_{\text{n}} \cdot S_{i}[t],
\end{equation}

\noindent where $\alpha_{\text{n}}$ is the glutamate clearance rate and $k_{\text{n}}$ is the glutamate release efficiency.

The released glutamate binds to metabotropic glutamate receptors on the astrocyte membrane and triggers the generation of a calcium pulse in the astrocyte \citep{Semyanov2020}. Traditionally, the process of glutamate-induced calcium transient generation in astrocytes is described by the widely used biophysical model proposed by Ullah et al.~\citep{ULLAH2006}. Such detailed models are computationally expensive for large networks and are unsuitable for hardware implementation. For the purposes of this work, we used a simplified phenomenological model that describes changes in the intracellular calcium concentration as a function of glutamate level and qualitatively reproduces the shape of the calcium transient in the Ullah model (Section~\ref{network1_4}). Therefore, we describe the intracellular calcium dynamics in astrocytes by a phenomenological second-order system (Figs.~\ref{fig1}G and~\ref{fig2}A):

\begin{equation} \label{eq:ca}
\begin{aligned}
\frac{d[\text{Ca}^{2+}]_i}{dt} &= y_i, \\
\frac{dy_i}{dt} &= \frac{J_{\text{neuro, i}} - 2\tau_{\text{Ca}} y_i - [\text{Ca}^{2+}]_i}{\tau_{\text{Ca}}^2},
\end{aligned}
\end{equation}

\noindent where $[\text{Ca}^{2+}]_i$ is the intracellular calcium concentration in the $i$-th astrocyte, $y_i$ is an auxiliary variable representing the rate of change of calcium concentration, $\tau_{\text{Ca}}$ characterizes the temporal dynamics, and $J_{\text{neuro, i}}$ represents the glutamate-induced stimulation:

\begin{equation}
J_{\text{neuro, i}} = 
\begin{cases}
A, & \text{if } G_i > G_{\text{thr}}; \\
0, & \text{otherwise}.
\end{cases}
\end{equation}

\subsubsection{Neuron-Astrocyte Interaction}\label{network1_4}

Experimental studies have shown that astrocytes are able to modulate neuronal excitability and synaptic transmission via astrocyte-released glutamate \citep{Santello2019}. More precisely, we consider the experimentally confirmed phenomenon that astrocytic Ca$^{2+}$ impulses trigger gliotransmitter release, which induces an increase in interneuron excitability \citep{Pabst2016}. In this work, for simplicity, we model this effect by astrocytic calcium activity-induced decrease in the interneuron firing threshold from Eq. (\ref{eq:LIF_excitatory}) as follows:

\begin{equation}
v_{thr, i} = \frac{v_{\text{thr}}^*}{1 + \eta_{\mathrm{astro}} \cdot H([\text{Ca}^{2+}]_i - [\text{Ca}^{2+}]_{\mathrm{thr}})},
\end{equation}

\noindent where $v_{\text{thr}}^*$ is the baseline firing threshold of interneurons, $\eta_{\mathrm{astro}}$ is the astrocytic influence factor, $H$ is the Heaviside step function, and $[\text{Ca}^{2+}]_{\mathrm{thr}}$ is the calcium threshold for gliotransmitter release. 

This astrocyte-mediated increase in interneuron excitability potentiates their inhibitory effect on motor neurons through the generation of inhibitory postsynaptic potentials (IPSPs). These IPSPs hyperpolarize the membrane potential of connected motor neurons (Fig.~\ref{fig1}H), suppressing their firing rate. As a result, the agent is unlikely to move to that location for the duration of the astrocytic modulation i.e. it can neither stay at the current grid cell nor come back to recently visited states.

Together with the STDP mechanism, this astrocytic modulation forms a complementary learning and memory system. While STDP reinforces pathways toward the goal by strengthening synapses between sequentially activated spatial cells, astrocytic activity provides short-term memory of recently visited positions. When the agent visits a location, neurotransmitter release activates the corresponding astrocyte, which then enhances inhibition of the associated motor neurons. This suppresses their firing rate, reducing the probability of revisits and preventing cyclic navigation. Thus, the dual mechanism supports efficient exploration: STDP guides the agent along reinforced paths to the goal, while astrocyte-mediated inhibition promotes exploration of novel areas by temporarily suppressing activity related to recent positions.

Model equations are integrated using the Euler method with a fixed timestep of 1 ms. A detailed listing of model parameters and their values can be found in Table~\ref{tab_1}. The code is available at \url{https://github.com/altergot/Neuromorphic_Agent_Navigation_Control_System_with_Short-Term_Memory}.

\begin{table*}[!tp]
\small
\caption{Spiking neuron-astrocyte network model parameters}
\label{tab_1}
\begin{tabular}{@{}>{\raggedright}p{0.13\textwidth} >{\raggedright}p{0.45\textwidth} >{\raggedright}p{0.14\textwidth} >{\raggedright\arraybackslash}p{0.14\textwidth}@{}}
\toprule
\textbf{Component} & \textbf{Parameter Description} & \multicolumn{2}{l}{\textbf{Value}} \\ 
\midrule

\multicolumn{4}{l}{\textbf{Network size}} \\
$N$ & Number of grid cells & \multicolumn{2}{l}{9 (100)} \\
$M×K$ & Grid dimensions & \multicolumn{2}{l}{$3×3$ ($10×10$)} \\
\midrule

\multicolumn{4}{l}{\textbf{Input stimulation}} \\
$p$ & Probability of generating a spike by an input neuron in each time step & \multicolumn{2}{l}{0.15} \\
$T_{\mathrm{stim}}$ & Stimulation and decision window duration & \multicolumn{2}{l}{1 s} \\
$\Delta t$ & Simulation time step & \multicolumn{2}{l}{1 ms} \\
\midrule

\multicolumn{4}{l}{\textbf{LIF neuron model}} \\
$\tau_m$ & Membrane time constant & \multicolumn{2}{l}{150 ms} \\
$v_{\mathrm{rest}}$ & Resting membrane potential & \multicolumn{2}{l}{0} \\
$v_{\mathrm{thr}}$ & Spike threshold potential (motor neurons) & \multicolumn{2}{l}{7}\\
$v_{\mathrm{thr}}^*$ & Spike threshold potential (interneuron) & \multicolumn{2}{l}{7} \\
$v_{\mathrm{reset}}$ & Reset potential after a spike & \multicolumn{2}{l}{0} \\
$t_{\mathrm{ref}}$ & Refractory period & \multicolumn{2}{l}{40 ms} \\
$R$ & Membrane resistance & \multicolumn{2}{l}{1 M$\Omega$} \\
$I_{\mathrm{bias}}$ & Interneuron bias current & \multicolumn{2}{l}{0.02} \\
\midrule

\multicolumn{4}{l}{\textbf{Synaptic connections}} \\
$w$ & Input $\to$ motor synaptic weights (plastic) & \multicolumn{2}{p{0.3\textwidth}}{Randomly selected from normal distribution ($\mu$ = 0.65, $\sigma$ = 0.1) and $\in$ [0.5, 0.8]} \\
$w_{I}$ & Input $\to$ interneuron synaptic weights (fixed) & \multicolumn{2}{l}{Identity matrix} \\
$A_+$ & STDP potentiation amplitude & \multicolumn{2}{l}{0.04} \\
$A_-$ & STDP depression amplitude & \multicolumn{2}{l}{0.03} \\
$\tau_+$ & STDP potentiation time constant & \multicolumn{2}{l}{8 ms} \\
$\tau_-$ & STDP depression time constant & \multicolumn{2}{l}{8 ms} \\
$w_{\mathrm{min}}$ & Minimum synaptic weight & \multicolumn{2}{l}{0.001} \\
$w_{\mathrm{max}}$ & Maximum synaptic weight & \multicolumn{2}{l}{1.0} \\
$k$ & Pathway reinforcement factor & \multicolumn{2}{l}{1.5} \\
\midrule

\multicolumn{4}{l}{\textbf{Astrocyte model}} \\
& & {\textbf{3×3}} & {\textbf{10×10}}  \\
$A$ & Calcium impulse amplitude parameter &10  &110\\
$\tau_{\mathrm{Ca}}$ & Calcium time constant & 2 s & 22 s\\
\midrule

\multicolumn{4}{l}{\textbf{Neuron-astrocyte interaction}} \\
$\alpha_{\mathrm{n}}$ & Neurotransmitter clearance rate &  \multicolumn{2}{l} {0.001 ms$^{-1}$} \\
$k_{\mathrm{n}}$ & Efficacy of neurotransmitter release & \multicolumn{2}{l} {0.2} \\
$G_{\mathrm{thr}}$ & Threshold concentration of neurotransmitter for astrocyte activation & \multicolumn{2}{l} {0.2} \\
$[\mathrm{Ca}^{2+}]_{\mathrm{thr}}$ & Threshold concentration of Ca$^{2+}$ for gliotransmitter release & \multicolumn{2}{l} {0.2} \\
$ \eta_{\mathrm{astro}}$ & Astrocytic influence factor on interneuron threshold & \multicolumn{2}{l} {5} \\
\bottomrule
\end{tabular}
\par\smallskip
\begin{minipage}{1\textwidth}\footnotesize
The parameters $A$ and $\tau_{\mathrm{Ca}}$ of the astrocyte model were selected to provide the required astrocytic calcium signal duration and, accordingly, the required astrocyte-induced short-term memory capacity.
\end{minipage}
\end{table*}

\subsection{Memristive Spiking Neural Network}\label{network2}

\begin{figure}
\centering
\includegraphics[width=1\columnwidth]{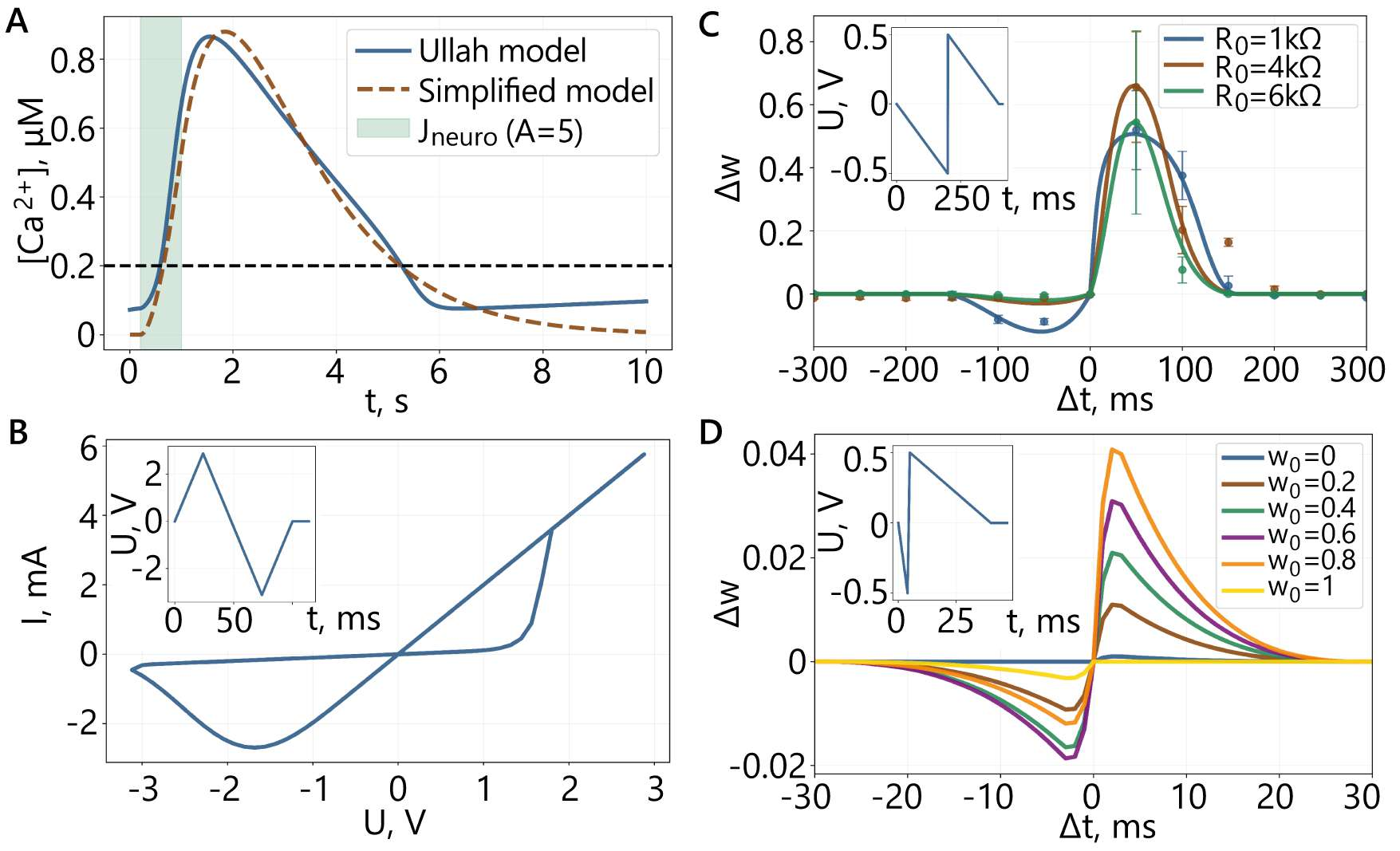}
\caption{
(A) Astrocytic calcium dynamics: detailed biophysical Ullah model (blue) and the phenomenological model used in this work (brown). (B) Approximation of the experimental STDP curves by the VTEAM model for different initial resistive states of the PPX-based memristor. The insert shows an example of a bi-triangular pulse used to obtain the curves. (C)  Current-voltage characteristic of the VTEAM model of the PPX-based memristor. The insert shows the pulse used to obtain the characteristic. (D) Memristive STDP window curves used in SNN simulations for 5 initial normalized weight values. The insert shows the spike pulse used in the SNN simulation. }
\label{fig2}
\end{figure}

To approach hardware implementation, we replaced the STDP model description with the Voltage Threshold Adaptive Memristor (VTEAM) model \citep{Kvatinsky2015} for synaptic plasticity. The VTEAM model was selected for its computational efficiency, generality, flexibility, and high accuracy in capturing memristive device behavior, with parameters fitted to experimental STDP characteristics of bipolar poly-para-xylylene (parylene) memristors \citep{Matsukatova2023} (Fig.~\ref{fig2}B-D). This model assumes negligible conductance changes within the voltage range bounded by device-specific thresholds. The dynamics of the effective tunnel barrier thickness are described by:

\begin{equation}
\label{eq:vteam1}
\begin{aligned}
\frac{dx(t)}{dt} &= K_{\text{off}} \left( \frac{v(t)}{v_{\text{off}}} - 1 \right)^{A_{\text{off}}} f_{\text{off}}(x), & v \leq v_{\text{off}} < 0; \\
\frac{dx(t)}{dt} &= 0, & v_{\text{off}} < v < v_{\text{on}};\\
\frac{dx(t)}{dt} &= K_{\text{on}} \left( \frac{v(t)}{v_{\text{on}}} - 1 \right)^{A_{\text{on}}} f_{\text{on}}(x), & 0 < v_{\text{on}} \leq v,
\end{aligned}
\end{equation}
where $K_{\text{off}} > 0$, $K_{\text{on}} < 0$, $A_{\text{off}}$, $A_{\text{on}}$, $a_{\text{off}}$, and $a_{\text{on}}$ are constants determined by fitting experimental I-V characteristics and STDP data; $v_{\text{off}}$ and $v_{\text{on}}$ are the threshold switching voltages to high-resistance and low-resistance states, respectively; and $f_{\text{off}}(x)$ and $f_{\text{on}}(x)$ are window functions accounting for decreased ion mobility near device boundaries:

\begin{equation}
\label{eq:vteam2}
\begin{aligned}
f_{\text{off}}(x) &= \exp\left(-\exp\left(\frac{x - a_{\text{off}}}{x_c}\right)\right), \\
f_{\text{on}}(x) &= \exp\left(-\exp\left(\frac{a_{\text{on}} - x}{x_{\text{con}}}\right)\right).
\end{aligned}
\end{equation}

The memristance is described by:

\begin{equation}
\label{eq:vteam3}
\begin{aligned}
R &= R_{\text{on}} \exp\left(\lambda \frac{x - x_{\text{on}}}{x_{\text{off}} - x_{\text{on}}}\right), \\
\lambda &= \ln\left(\frac{R_{\text{off}}}{R_{\text{on}}}\right).
\end{aligned}
\end{equation}

A detailed listing of VTEAM model parameters and their values can be found in Table~\ref{tab_2}. This specific type of parylene memristors is used here only as an example. The VTEAM model has been shown to be applicable to a wide range of different types of bipolar memristors \citep{Kvatinsky2015, Matsukatova2022}, so that, conceptually, the suggested SNAN system can be realized based on different types of memristors with STDP-like plasticity. STDP curves for the memristor model were measured for different initial resistive states with a step size of 1$\%$ of the available conductance range. For each initial state, a voltage-time dependency was applied to the model input, representing the subtraction of two bipolar triangular pulses with different time delays relative to each other. Each individual pulse was unable to change the memristor's resistive state, as its amplitude did not reach the threshold values. Thus, the synaptic connections implemented as memristive elements exhibited plasticity governed by the STDP rule (Fig.~\ref{fig2}D). 

\begin{table}[!htbp]
\centering
\caption{VTEAM model parameters}
\label{tab_2}
\begin{tabular}{p{1.5cm}p{3.5cm}c}
\toprule
\textbf{Component} & \textbf{Parameter} & \textbf{Value} \\ 
\midrule
$v_{\mathrm{on}}$ & SET threshold voltage & 0.6 V \\
$v_{\mathrm{off}}$ & RESET threshold voltage & -0.6 V \\
$R_{\mathrm{on}}$ & Low resistance state & 500 $\Omega$ \\
$R_{\mathrm{off}}$ & High resistance state & 25 k$\Omega$ \\
$K_{\mathrm{on}}$ & SET rate constant & -3.96 $\times$ 10$^{-8}$ m/s \\
$K_{\mathrm{off}}$ & RESET rate constant & 1.054 $\times$ 10$^{-8}$ m/s \\
$A_{\mathrm{on}}$ & SET exponent & 2.357 \\
$A_{\mathrm{off}}$ & RESET exponent & 2.022 \\
$a_{\mathrm{on}}$ & SET window parameter & 36.8 nm \\
$a_{\mathrm{off}}$ & RESET window parameter & 0.12 nm \\
$x_c$ & RESET window scaling factor & 40 nm \\
$x_{\mathrm{con}}$ & SET window scaling factor & 38 nm \\
\bottomrule
\end{tabular}
\end{table}

\section{Results}\label{sec_results}
\subsection{Minimal-Sized Grid Environment}\label{sec_results1}

We first evaluated the performance of the navigating agent using a neurophysiologically inspired short-term memory model implemented by SNAN in an initially unfamiliar, partially observable environment: a small 3 $\times$ 3 square grid. In this case, we aim to demonstrate the impact of astrocyte-induced short-term memory on the efficiency of agent navigation. Initially, we excluded the astrocytic influence from the system. We tested our agent in 1000 independent trials. For each trial, a new initial agent state and goal are randomly set. As a metric of the agent’s performance during a trial, we use the number of steps required for the agent to reach the goal. The trial ends when the agent reaches the goal state or when the agent has made 100 steps. The agent receives no observations about the whole environment; it observes only its current position (cell) and is unable to move beyond the grid boundaries. Unless otherwise specified, the synaptic weights were re-initialized before each trial. To assess the contribution of synaptic plasticity within a single trial, three variants of the network were compared: without STDP, with additive STDP (Eq.~\ref{eq:stdp}), and with memristive STDP (Section~\ref{network2}). The effect of accumulating STDP-based weight changes over a sequence of trials is analyzed separately in Section~\ref{sec_results6}.

\begin{figure}
\centering
\includegraphics[width=1\columnwidth]{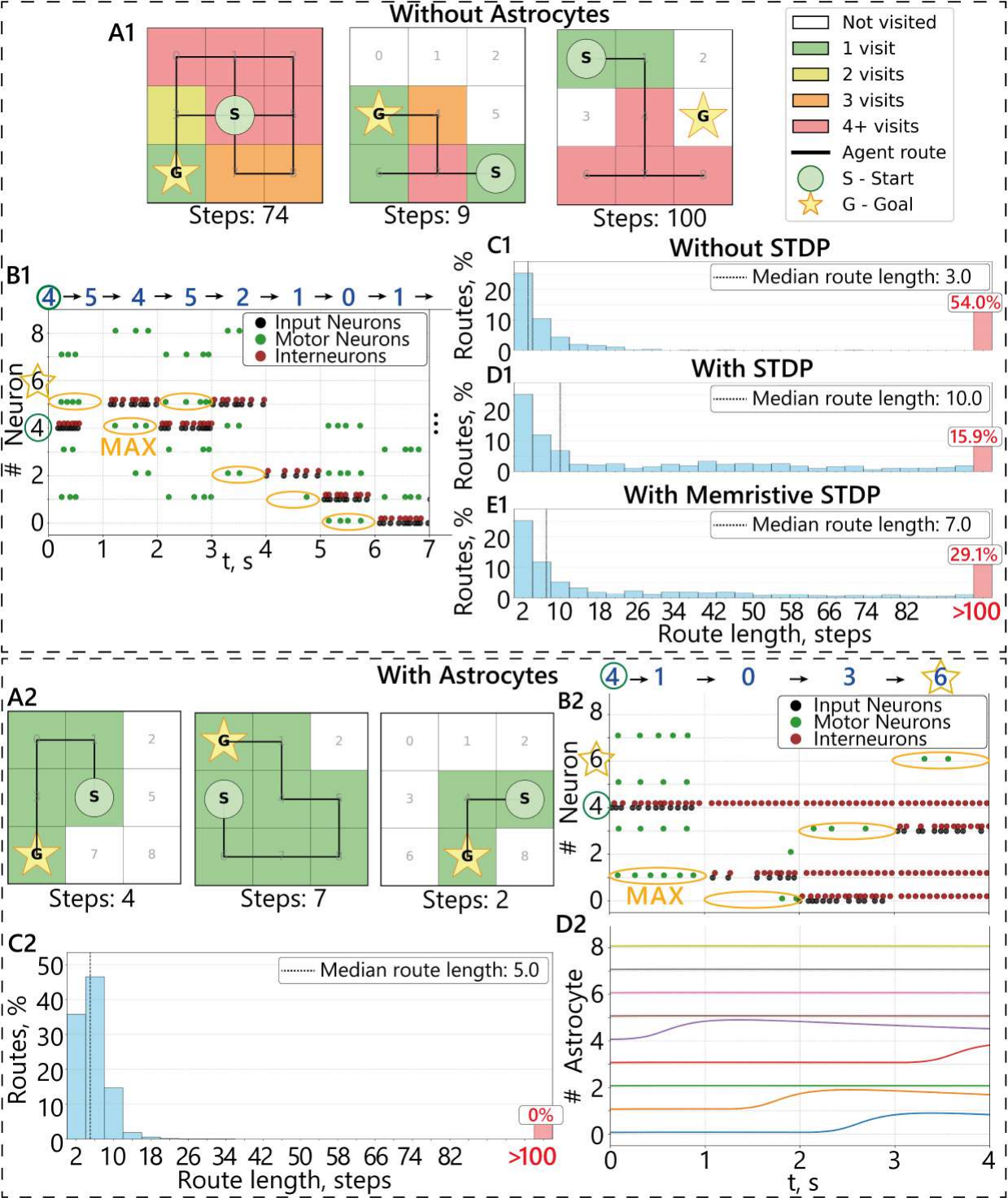}
\caption{Comparative analysis of agent navigation performance in a minimal-sized environment ($N = 9$). The short-term memory agent model was implemented by SNAN either without astrocytes (A1–E1) or with them (A2–D2). (A1, A2) Representative agent trajectories. (B1, B2) Raster plots of neural activity during agent navigation: input neuron (black), output motor neuron (green), and interneuron (red). (C1--E1) Distributions of route lengths from 1000 trials for the network without astrocytes: (C1) without STDP, (D1) with additive STDP, and (E1) with memristive STDP. (C2) Distribution of route lengths for the network with astrocyte-induced short-term memory and memristive STDP. The maximum route length was 100 steps, and the rightmost bar shows the fraction of failed trials. (D2) Intracellular calcium dynamics in the astrocytic network during agent navigation.}
\label{fig3}
\end{figure}

Fig.~\ref{fig3}A1 presents representative agent trajectories, revealing frequent cyclic patterns and suboptimal pathfinding behavior. The corresponding raster plot in Fig.~\ref{fig3}B1 illustrates the neural activity dynamics during agent navigation, showing input neuron stimulation (black), output motor neuron responses (green), and interneuron activity (red).

In the absence of short-term memory for recent steps, implemented in our system via neuron–astrocyte interactions, statistical analysis of 1000 independent trials (Fig.~\ref{fig3}C1--E1) indicates limited navigation efficiency. Without STDP, the agent reached the goal in only 46.0$\%$ of trials (Fig.~\ref{fig3}C1). In this case, successful trajectories were very short (median 3 steps), because the goal was generally reached only when it was located close to the starting point, before the agent became trapped in a cycle. The addition of synaptic plasticity increases reliability: the agent completed 84.1$\%$ of trajectories with additive STDP (median 10 steps, Fig.~\ref{fig3}D1) and 70.9$\%$ with memristive STDP (median 7 steps, Fig.~\ref{fig3}E1). This occurs because, for long trajectories, STDP equalizes the weights of connections that were initially unequal, thereby making the agent’s cell-to-cell transitions equiprobable and allowing the agent to escape cyclic behavior. Nevertheless, 15.9--29.1$\%$ of trials still failed within the 100-step limit due to cyclic behavior.

The inclusion of astrocytic modulation of neural network dynamics in our agent substantially improves navigation performance. Fig.~\ref{fig3}A2 shows the agent's trajectories, which are characterized by direct paths to the goal with minimal detours. The corresponding raster plot in Fig.~\ref{fig3}B2 shows the dynamics of neural activity during the agent's navigation. Interneuron activity (red) exhibits temporal patterning correlated with calcium activity in the associated astrocytes (Fig.~\ref{fig3}D2). For a given set of parameter values, the astrocytic influence on the excitability of inhibitory neurons lasts 10 seconds, resulting in a short-term memory capacity of 10 steps for the agent. In this case, the agent reached the goal in 100$\%$ of trials regardless of whether STDP was absent, additive, or memristive, with a median route length of 5 steps (Fig.~\ref{fig3}C2). Considering only the routes successfully completed by the agent using the navigation system with memristive STDP, both with and without astrocytes ($n=709$), the addition of short-term memory reduced the median route length from 7 to 4 steps (Wilcoxon signed-rank test, $p<0.001$). 

Thus, additive and memristive STDP produced similar results. Since this work focuses on hardware implementation and compatibility with memristive devices, all subsequent results on larger-scale environment navigation tasks are reported for the network with memristive STDP.

\subsection{Scalability to Larger Environments} \label{sec_results2}
Next, we tested our agent in a larger environment, considering the case of a 10×10 square grid ($N$=100). In this case, the agent was allowed a maximum of 1000 steps per trial, and 1000 independent trials were considered. For this environment, we again compared the performance of the agent's bioinspired navigation system under conditions with and without astrocytic influence. 

\begin{figure}
\centering
\includegraphics[width=1\columnwidth]{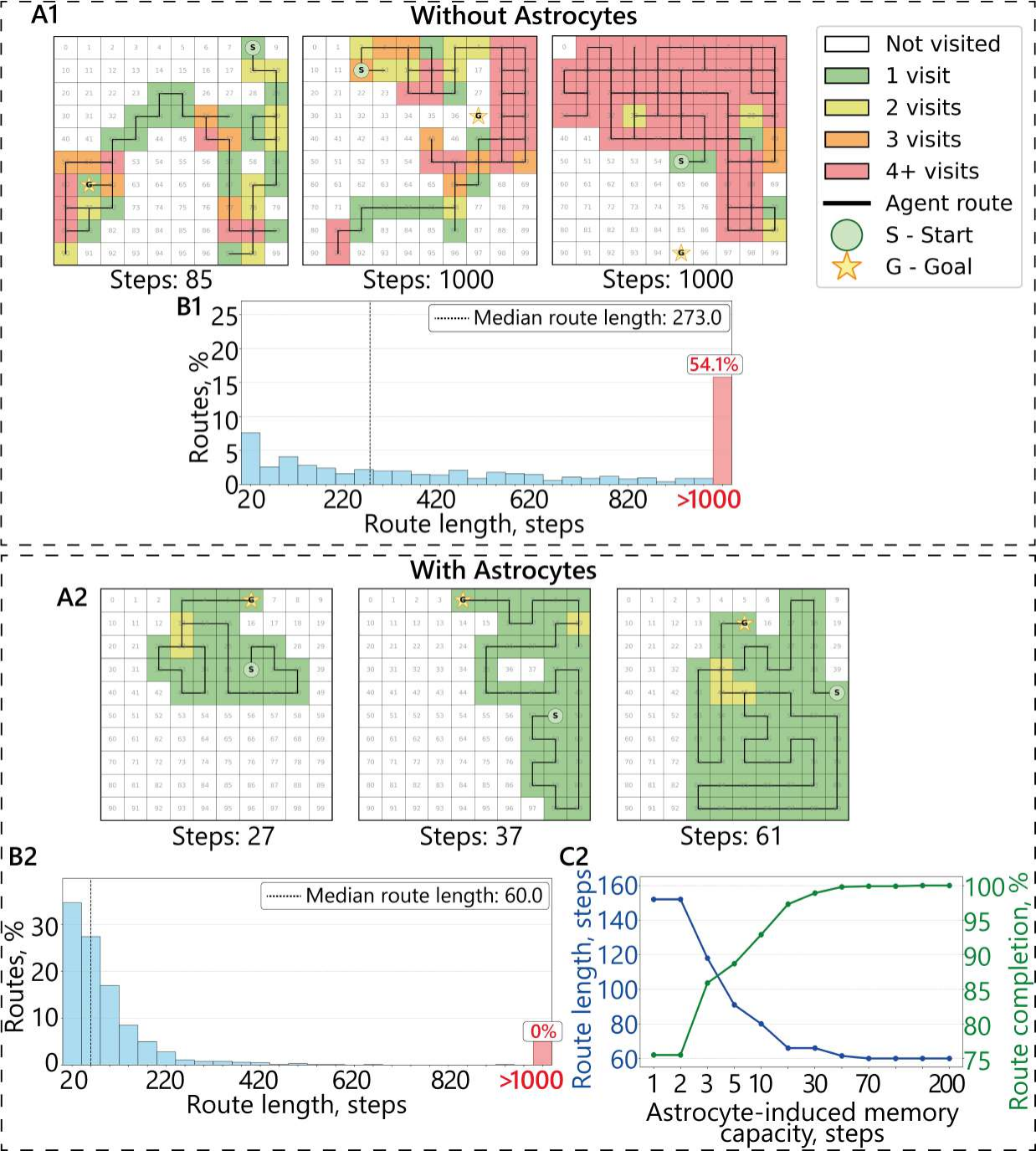}
\caption{Comparative analysis of agent navigation performance in a large-scale grid environment ($N = 100$). The short-term memory agent model was implemented by SNAN either without astrocytes (A1–B1) or with them (A2–C2). (A1, A2) Representative agent trajectories. (B1, B2) Route length histograms from 1000 trials. Without astrocytes: 45.9$\%$ success rate (median length: 273 steps). With astrocytes: 99.9$\%$ success rate (median length: 60 steps). (C2) Median route length and fraction of completed routes as functions of calcium event duration. For (A2) and (B2), $\tau_{\mathrm{Ca}}=22, A=110$.}
\label{fig4}
\end{figure}

In the absence of astrocytic regulation, agent navigation was markedly inefficient: representative trajectories exhibited extensive cyclic behavior (Fig.~\ref{fig4}A1), with only 45.9$\%$ of trials successfully reaching the goal within the 1000-step limit and a median path length of 273 steps among successful attempts (Fig.~\ref{fig4}B1). In contrast, implementation of the memristive SNAN yielded substantially improved performance. Trajectories became more direct with reduced cycling (Fig.~\ref{fig4}A2), achieving a 99.9$\%$ success rate and a markedly reduced median path length of 60 steps (Fig.~\ref{fig4}B2). When comparing the contribution of astrocytic modulation for successfully completed tasks only ($n=459$), the addition of astrocytes reduced the median route length by a factor of 5.8, from 273 to 47 steps (Wilcoxon signed-rank test, $p<0.001$). 

Importantly, as the environment size increased and potential route lengths became longer, we had to increase the memory capacity and, accordingly, the duration of astrocytic modulation to 100 steps. This effect, as described previously (Section~\ref{network1_4}), depends on the duration of the calcium transient in the astrocyte, which is determined by the parameters $\tau_{\mathrm{Ca}}, A$ ($\tau_{\mathrm{Ca}}=22 s, A=110$). 
We investigated the relationship between route length and the astrocyte-induced short-term memory capacity, revealing optimal ranges of astrocytic modulation duration for large-scale navigation. Fig.~\ref{fig4}C2 (blue) shows that the median route length decreases with increasing memory capacity and reaches a plateau of about 60 steps for the agent at memory capacities of 50 steps or more. Fig.~\ref{fig4}C2 (green) illustrates that, as memory capacity increases, the proportion of successfully completed routes increases rapidly, from 75.5$\%$ at a memory capacity of 1 step to 98.9$\%$ at 30 steps and 99.9$\%$ at 60 steps, reaching 100$\%$ for calcium event durations of 150 steps or longer. This parameter-sensitivity analysis establishes short-term memory capacity — and, consequently, the calcium event duration — as a key regulatory mechanism in large-scale navigation tasks.

\subsection{Navigation in Mazes}\label{sec_results3}

\begin{figure}
\centering
\includegraphics[width=1\columnwidth]{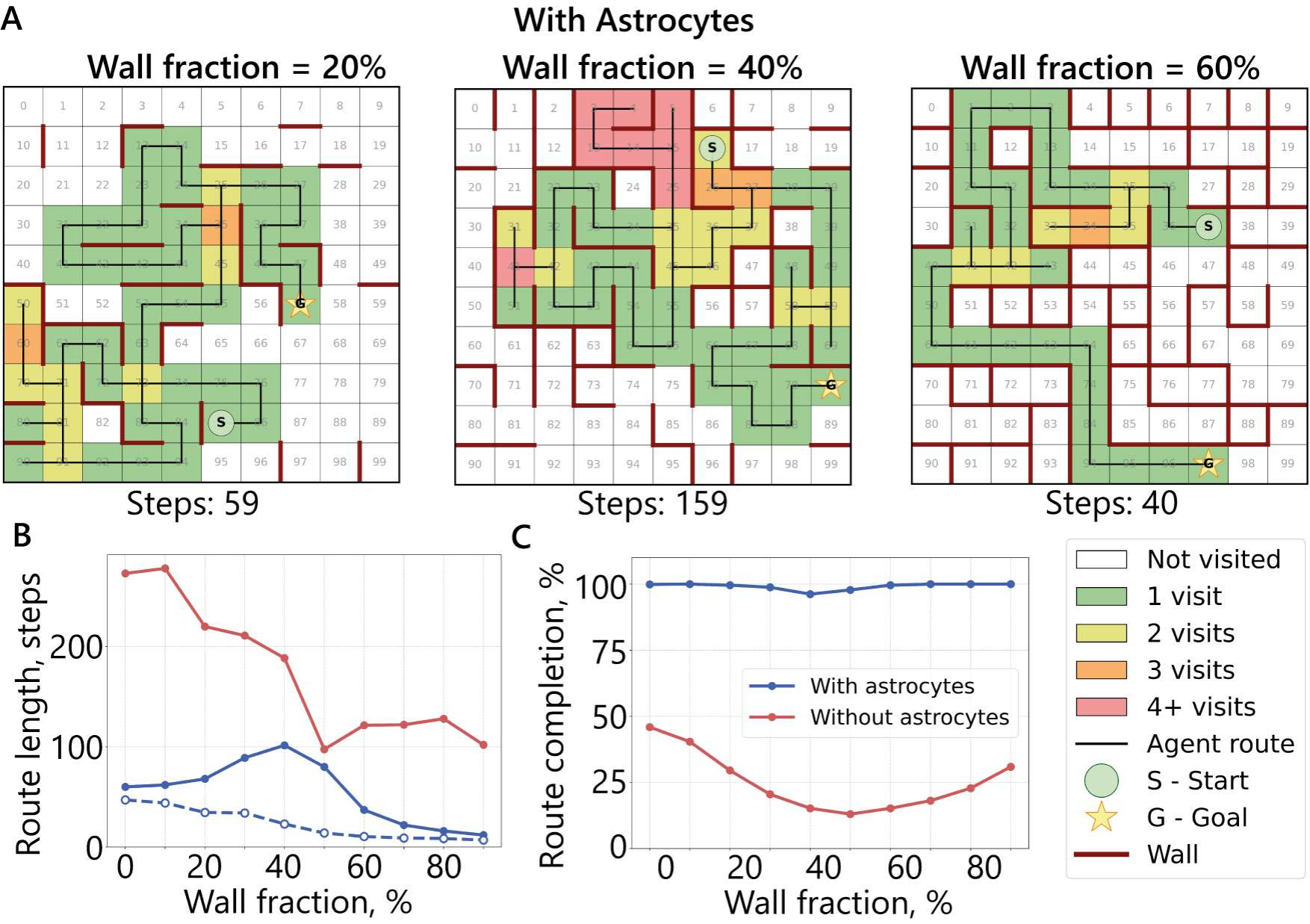}
\caption{Agent navigation performance in mazes.
(A) Representative agent trajectories controlled by SNAN in grid mazes with wall fractions of 20$\%$, 40$\%$, and 60$\%$. (B) Median length of completed routes as a function of wall fraction for the agent with (solid blue) and without (red) astrocyte-induced short-term memory. The dashed blue line shows the median route length of the agent with astrocytes calculated only over the task instances (same start, goal and maze) that were also completed by the agent without astrocytes, which removes the selection bias towards easy tasks in the astrocyte-free condition. (C) Fraction of completed routes as a function of wall fraction for the agent with (blue) and without (red) astrocyte-induced short-term memory. 1000 trials per wall fraction, memristive STDP, $\tau_{\mathrm{Ca}}=22, A=110$.}
\label{fig5}
\end{figure}

To further increase the complexity of the agent’s navigation task, we introduced obstacles into the grid environment. As a measure of maze complexity, we defined a parameter as the ratio of the total number of obstacle walls (i.e., walls between adjacent grid cells) to the total number of possible inter-cell walls in the entire grid. In this setup, the agent was allowed up to 1000 steps per trial, and 1000 independent trials were conducted. For each trial, obstacles were randomly generated, provided that the goal remained reachable from the starting position. We will consider two types of mazes: mazes with a static obstacle topology and mazes with a dynamic topology, in which the positions of the obstacles change as the agent navigates the maze (Sec.~\ref{sec_dyn_cmp}). Let us first consider the case of mazes with a static obstacle topology. Fig.~\ref{fig5}A shows representative agent trajectories controlled by SNAN in grid mazes with wall fractions of 20$\%$, 40$\%$, and 60$\%$.

Comparative analysis of navigation performance as a function of wall fraction (Fig.~\ref{fig5}B--C) demonstrates that astrocyte-induced short-term memory substantially enhances the reliability and efficiency of the agent’s navigation system. The system with astrocytic modulation exhibited a non-monotonic dependence of median route length on obstacle density (Fig.~\ref{fig5}B, solid blue line), with a peak at intermediate densities (40$\%$ walls: 101.5 steps) and improved performance at higher densities due to spatial constraints. Success rates remained consistently high across all obstacle densities (96.2--100$\%$; Fig.~\ref{fig5}C). In contrast, the agent without astrocytes completed only 13.0--45.9$\%$ of routes across all densities (Fig.~\ref{fig5}C), with the median length of its completed routes ranging from 97.5 to 278 steps (Fig.~\ref{fig5}B, red line).

A direct comparison of these medians is, however, biased, because the agent without astrocytes succeeds predominantly in the easier task instances (e.g., when the goal is located close to the start), whereas the agent with astrocytes also completes the more difficult ones. To remove this selection bias, the dashed blue line in Fig.~\ref{fig5}B shows the median route length of the agent with astrocytes computed only over the task instances (identical start, goal and maze) that were also completed by the agent without astrocytes. On this common subset, the routes of the agent with astrocytes were 5.8--15 times shorter at every wall fraction (e.g., 23 versus 188.5 steps at 40$\%$ walls).

\subsection{Multi-goal Navigation in Mazes}\label{sec_results4}

Then, we increased the complexity of the task within the same grid maze environment by requiring the agent to visit three different goal locations. In this task, the agent was allowed a maximum of 5000 steps per trial, with 1000 independent trials conducted for each wall fraction. The initial point and the goal positions were randomly selected for each trial, and the order in which the goals were visited was not prescribed.

\begin{figure}
\centering
\includegraphics[width=1\columnwidth]{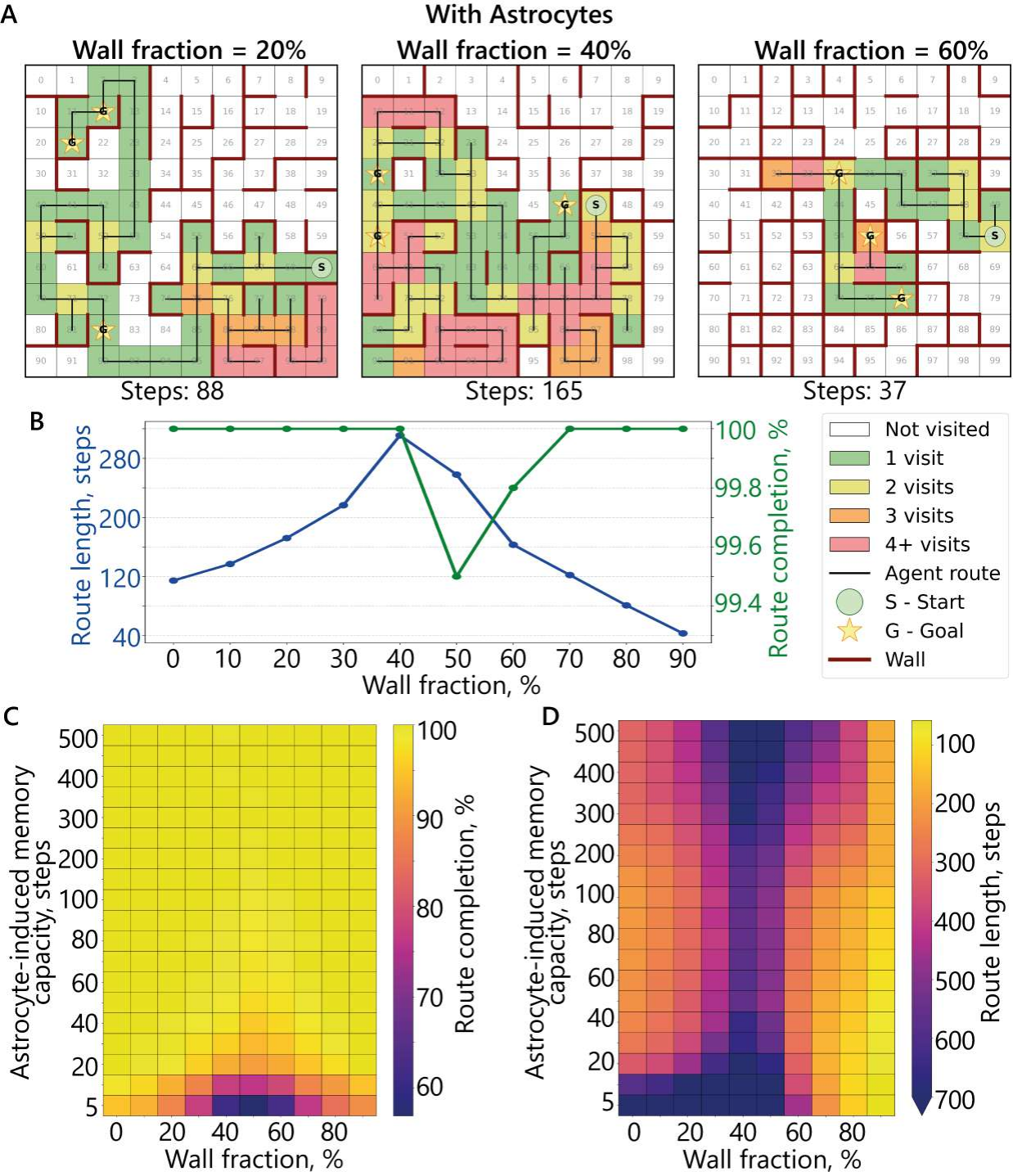}
\caption{Multi-goal navigation performance in mazes.
(A) Representative multi-goal agent trajectories controlled by bioinspired SNAN with short-term memory in mazes with 20$\%$, 40$\%$, and 60$\%$ wall fraction.
(B) Median route length and fraction of completed routes as functions of obstacle density for multi-goal navigation (1000 trials per configuration). For (A) and (B), $\tau_{\mathrm{Ca}}=22, A=110$. 
(C-D) Performance of the agent in multi-goal navigation tasks requiring a return to the start. (C) Two-parameter diagram mapping the dependence of the success rate on the duration of astrocyte-induced modulation of neural network dynamics for different obstacle densities in the maze.
(D) Two-parameter diagram mapping the dependence of the median route length on the duration of astrocyte-induced modulation of neural network dynamics for different obstacle densities in the maze. }
\label{fig6}
\end{figure}

Fig.~\ref{fig6}A shows representative multi-goal agent trajectories controlled by SNAN in mazes with 20$\%$, 40$\%$, and 60$\%$ wall fractions. We investigated the dependence of the median route length and the success rate on the obstacle density within the maze (Fig.~\ref{fig6}). The agent with the bioinspired navigation system, possessing short-term memory, demonstrated exceptional reliability, with success rates of 99.5--100$\%$ across all obstacle densities tested (Fig.~\ref{fig6}B, green). The agent achieved perfect completion rates (100$\%$) in environments with 0--40$\%$ and 70--90$\%$ wall coverage, with only minor performance variations at intermediate densities. The median route length exhibited a bell-shaped dependence on obstacle density, with the maximum path length observed at intermediate obstacle densities (40$\%$ walls: 311 steps). The median route length decreased significantly at higher obstacle densities (80$\%$ walls: 81 steps; 90$\%$ walls: 43 steps), indicating more constrained but efficient navigation in densely obstructed environments (Fig.~\ref{fig6}B, blue). In the absence of short-term memory, the agent mostly failed to complete this task: at the tested wall fractions of 0--90$\%$, the astrocyte-free network failed to visit all three goals within the 5000-step limit in 63.3--99.2$\%$ of attempts, with failure rates exceeding 98$\%$ at intermediate and high obstacle densities (40--80$\%$ wall coverage).

\textit{Multi-goal Navigation in Mazes with Return to Start}

Subsequently, the task complexity was further increased by requiring the agent to return to the starting point after visiting all three targets (Fig.~\ref{fig6}C--D). The agent navigated by the system without short-term memory rarely solved this task: it completed 21.9$\%$ of routes in the open grid, and only 0--0.1$\%$ at wall fractions of 40--80$\%$. In the case where the agent is controlled by the SNAN, navigation efficiency depends on the capacity of short-term memory, which is determined by the duration of astrocytic activity, as observed in the previous experiment (Fig.~\ref{fig4}C2). To investigate this effect, we constructed a two-parameter diagram mapping the dependence of the success rate on the astrocyte-induced short-term memory capacity for different obstacle densities in the maze (Fig.~\ref{fig6}C). As expected, limited memory capacity (5--10 steps) resulted in markedly higher failure rates, particularly in moderately complex environments (40--60$\%$ wall density, 56.9--78.8$\%$ of completed routes), suggesting that brief astrocytic modulation periods are insufficient for effective cycle prevention. As the astrocyte-induced memory capacity increased to 50 steps or more, the system achieved high success rates (96.9--100$\%$) across all obstacle densities, demonstrating that sustained astrocytic modulation provides the necessary temporal window for efficient obstacle avoidance. Investigating the dependence of route length on the same parameters (Fig.~\ref{fig6}D), we found that, over the memory-capacity range of 30–100 steps, the median route length for different obstacle densities did not depend monotonically on memory capacity and exhibited a bell-shaped curve similar to that shown in Fig.~\ref{fig6}B. Note that in this range of parameter values, the median route length is several times greater than the memory capacity, indicating that the system needs to remember only part of the path rather than the entire path. When the memory capacity becomes large, the agent’s route length increases, because astrocytes suppress almost all recently visited cells, including the corridors that must be traversed again to reach the remaining goals or to return to the start, and the agent’s movement approaches a random walk. The optimum was observed at an astrocyte-induced memory capacity of about 40–200 steps, where the system maintained high success rates while minimizing route lengths.

\subsection{Agent Route Learning Capability via STDP and Astrocytic Modulation}\label{sec_results6}

\begin{figure}
\centering
\includegraphics[width=1\columnwidth]{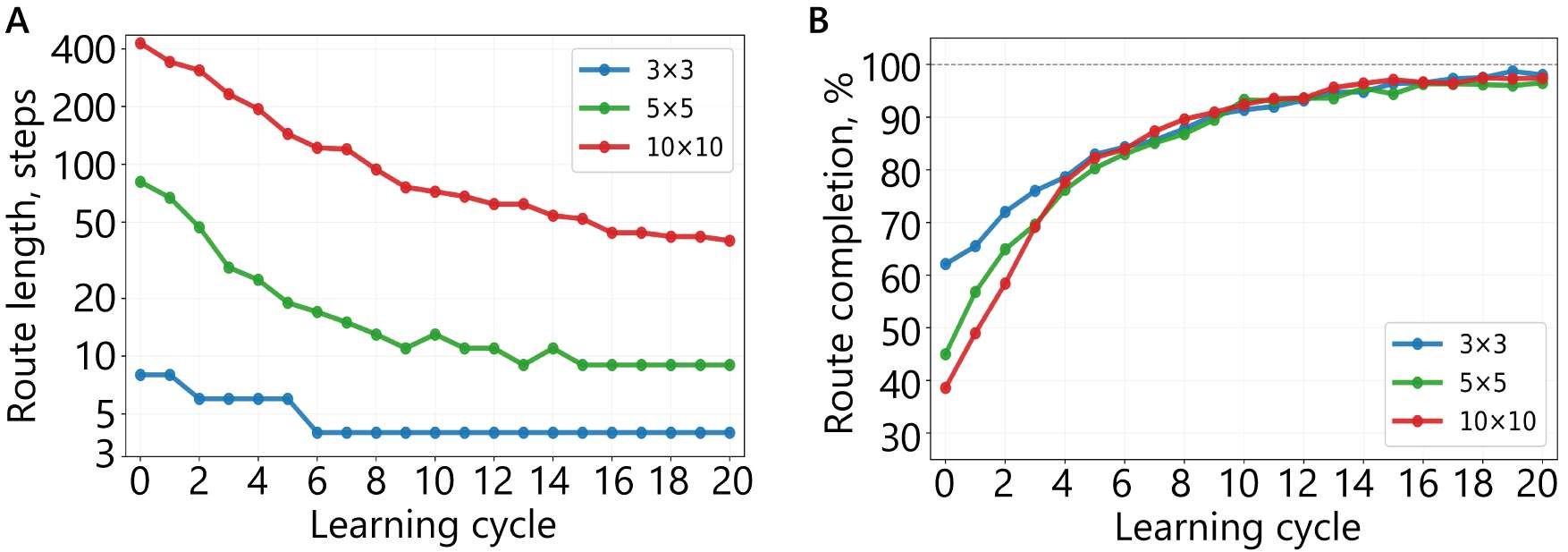}
\caption{Memristive STDP route learning performance across different environment sizes. 
(A) Median route length (steps) during test phases without astrocyte influence across 20 learning cycles for 3×3 (blue), 5×5 (green), and 10×10 (red) grids. Note the logarithmic scale. 
(B) Fraction of completed routes during the same test phases. Learning cycle 0 corresponds to initial performance before training. Maximum allowed steps: 100 (3×3), 300 (5×5), and 1000 (10×10).}
\label{fig7}
\end{figure}

To complete our evaluation of the agent navigation control capabilities using the bioinspired memristive neuron-astrocyte network with short-term memory, we examined the agent’s ability to learn and improve navigation routes through repeated experience. The experimental session consisted of 20 consecutive cycles of agent navigation within the same grid environment, without reinitializing the network's synaptic weights at the beginning of each cycle. During each cycle, while the agent was navigating to the goal, the synaptic weights of the SNAN were updated according to a synaptic plasticity rule with astrocytic modulation. Navigation tasks were considered in square grid environments of varying sizes (3×3, 5×5, and 10×10) containing no obstacles, analogous to the tasks described in Sections~\ref{sec_results1} and~\ref{sec_results2}. At the beginning of each experimental session, a baseline test of the agent's navigation performance was conducted without astrocytic influence, or short-term memory (Learning cycle 0 in Fig.~\ref{fig7}A). To assess the dynamics of the agent's navigation performance as it learned the route over the course of the session, we tested the agent after each of the 20 cycles using the synaptic weight values from the preceding cycle and in the absence of astrocytes. For each grid environment size, we conducted 1000 independent experimental sessions with random initial synaptic weights and a fixed start--goal pair. Performance was evaluated using the route length and the fraction of completed routes—i.e., the proportion of trials in which the agent reached the goal within the maximum allowed number of steps (100 for the 3×3, 300 for the 5×5, and 1000 for the 10×10 grids).

Fig.~\ref{fig7}A shows the median route length across test phases. For the 10×10 environment, route length decreased from 428 steps (cycle 0, before any learning) to 40 steps (cycle 20, after 20 learning cycles). The 5×5 environment improved from 81 to 9 steps, and the 3×3 environment from 8 to 4 steps. Fig.~\ref{fig7}B displays the corresponding fraction of completed routes. The 10×10 environment showed the most significant improvement, increasing from 38.6\% to 97.4\%. The 5×5 environment improved from 45.0\% to 96.5\%, and the 3×3 environment from 62.1\% to 98.0\%. These results demonstrate that memristive STDP enables effective route learning. During training phases, astrocyte-mediated inhibition prevents cyclic navigation and promotes exploration of alternative paths, generating diverse but relevant experiences for synaptic plasticity. STDP then reinforces successful routes through accumulation of weight updates, leading to progressive performance improvement across training cycles.

\subsection{Comparison with Classical Navigation Strategies}\label{sec_results7}

Finally, we compared the proposed SNAN-based navigation system with classical navigation strategies in a partially observable grid environment. All agents operated under the same conditions as those described in Section~2.1: each agent observed only its current position and the admissible transitions from that position, with no information about the goal location. We used seven baseline strategies. The random walk selects a random admissible move \citep{Lovasz1993}. The random mouse also moves randomly but does not return to the previous position unless no other move is available, i.e., it performs a non-backtracking random walk \citep{Alon2007}. The wall follower follows the left-hand rule \citep{Lumelsky1987}, whereas the Pledge algorithm combines wall following with a preferred direction and a turn counter \citep{Abelson1980}. Tr\'emaux's algorithm marks every traversed transition, prefers unmarked transitions, and never uses a transition marked twice \citep{Lucas1882,Even2011}. Tarry's algorithm never traverses a transition twice in the same direction and uses the transition through which it first entered a cell only as a last resort \citep{Tarry1895}. These two algorithms traverse every transition of a finite connected maze at most twice and are therefore guaranteed to reach every reachable goal \citep{Tarry1895,Even2011}. Under the present observation model, they constitute the classical exact solutions to the maze-exploration problem. As the seventh baseline, we used tabular Q-learning \citep{Watkins1989,Watkins1992} with an $\varepsilon$-greedy policy ($\varepsilon=0.1$), a learning rate of $0.5$, a discount factor of $0.95$, and a reward of $-1$ per step and $+10$ upon reaching a goal; it was trained from scratch for each task unless stated otherwise. Agents based on recurrent networks or meta-reinforcement learning were excluded because they require pre-training on large sets of environments \citep{Duan2016,Zintgraf2020}, whereas all strategies considered here start without prior knowledge of the environment. All agents were tested on exactly the same task instances (start, goals, and maze). For each pair of agents, route lengths were compared only on tasks completed by both agents, using the two-sided Wilcoxon signed-rank test \citep{Wilcoxon1945}. All comparisons were performed in the $10\times10$ grid.

\begin{table*}[!tp]
\caption{Comparison of the neuromorphic navigation system with classical navigation strategies in the $10\times10$ grid on identical task instances (1000 trials per row).}
\label{tab_compare}
\newlength{\memw}\setlength{\memw}{0.088\textwidth}
\scriptsize
\setlength{\tabcolsep}{3pt}
\begin{tabular*}{\textwidth}{@{\extracolsep{\fill}}lcccccccc@{}}
\toprule
Task, wall fraction & Our SNAN & RW & RM & WF & Pledge & Tr\'emaux & Tarry & Q \\
\midrule
\multicolumn{9}{@{}l}{\textit{Memory requirements within a trial}} \\
Stored information & \parbox[t]{\memw}{\centering Astrocyte activity (Ca$^{2+}$ level) per position ($n$)} & \parbox[t]{\memw}{\centering none} & \parbox[t]{\memw}{\centering Arrival direction} & \parbox[t]{\memw}{\centering Heading, wall-following flag } & \parbox[t]{\memw}{\centering Heading, preferred heading, wall-following flag, turn counter } & \parbox[t]{\memw}{\centering Transition marks, visited flags, arrival direction, re-entry flag ($E+n+2$)} & \parbox[t]{\memw}{\centering Directed-transition flags, first-entry directions ($2E+n$)} & \parbox[t]{\memw}{\centering State--action values ($2E$)} \\[1pt]
Positions, open 10×10 grid & 100 & 0 & 1 & 2 & 4 & 282 & 460 & 360 \\
\midrule
\multicolumn{9}{@{}l}{\textit{Static mazes}} \\
1 goal, 0\% & \cellcolor{bestgreen}\textbf{99.9/60} & 97.2/146 & 99.8/97 & \textcolor{lowgray}{8.7/5} & \textcolor{lowgray}{38.4/20} & 100.0/86 & 100.0/75 & 100.0/132 \\
1 goal, 20\% & \cellcolor{bestgreen}\textbf{99.6/68} & 91.9/189 & 98.9/99 & \textcolor{lowgray}{17.7/24} & \textcolor{lowgray}{40.3/24} & 100.0/80 & 100.0/79 & 100.0/154.5 \\
1 goal, 40\% & 96.2/101.5 & \textcolor{lowgray}{78.2/252} & 97.1/116 & \textcolor{lowgray}{63.4/48} & \textcolor{lowgray}{53.2/43} & \cellcolor{bestgreen}\textbf{100.0/68} & 100.0/77 & 100.0/167.5 \\
1 goal, 60\% & 99.6/37 & \textcolor{lowgray}{87.4/189.5} & 99.9/53 & \cellcolor{bestgreen}\textbf{96.5/27} & \textcolor{lowgray}{60.5/26} & 100.0/29 & 100.0/30 & 100.0/103.5 \\
3 goals, 0\% & \cellcolor{bestgreen}\textbf{100.0/114.5} & 100.0/373 & 100.0/215.5 & \textcolor{lowgray}{0.4/30} & \textcolor{lowgray}{4.3/31} & 100.0/173 & 100.0/154 & 100.0/266.5 \\
3 goals, 20\% & 100.0/172 & 100.0/501 & 100.0/242.5 & \textcolor{lowgray}{4.7/49} & \textcolor{lowgray}{9.4/44.5} & 100.0/159 & \cellcolor{bestgreen}\textbf{100.0/152} & 100.0/316 \\
3 goals, 40\% & 100.0/311 & 96.9/970 & 100.0/299 & \textcolor{lowgray}{47.9/88} & \textcolor{lowgray}{34.2/94} & 100.0/136 & \cellcolor{bestgreen}\textbf{100.0/135} & 100.0/386 \\
3 goals, 60\% & 99.8/163 & 99.2/842.5 & 100.0/179 & \cellcolor{bestgreen}\textbf{97.4/62} & \textcolor{lowgray}{52.9/66} & 100.0/64 & 100.0/65 & 100.0/246.5 \\
3 goals + return, 0\% & 100.0/208 & 100.0/618 & 100.0/344 & \textcolor{lowgray}{0.4/36} & \textcolor{lowgray}{1.4/36} & 100.0/234 & \cellcolor{bestgreen}\textbf{100.0/186} & 100.0/484 \\
3 goals + return, 20\% & 100.0/298 & 100.0/834 & 100.0/407 & \textcolor{lowgray}{3.9/64} & \textcolor{lowgray}{5.0/64} & 100.0/206 & \cellcolor{bestgreen}\textbf{100.0/186} & 100.0/618 \\
3 goals + return, 40\% & 99.7/576 & 92.2/1666 & 100.0/536 & \textcolor{lowgray}{47.0/122} & \textcolor{lowgray}{30.7/130} & \cellcolor{bestgreen}\textbf{100.0/158} & \cellcolor{bestgreen}\textbf{100.0/158} & 100.0/758 \\
3 goals + return, 60\% & 99.5/298 & 95.2/1438 & 100.0/307 & 97.4/90 & \textcolor{lowgray}{52.2/96} & \cellcolor{bestgreen}\textbf{100.0/80} & \cellcolor{bestgreen}\textbf{100.0/80} & 100.0/441 \\
\midrule
\multicolumn{9}{@{}l}{\textit{Dynamic mazes}} \\
1 goal, 20\% & \cellcolor{bestgreen}\textbf{100.0/68} & 96.6/178.5 & 99.6/93 & 97.1/158 & 96.4/103.5 & 100.0/86 & 99.9/81 & 96.1/139 \\
1 goal, 40\% & \cellcolor{bestgreen}\textbf{100.0/70} & 96.2/164.5 & 99.9/91 & 99.9/96 & 99.8/100 & 100.0/80 & 99.9/79 & \textcolor{lowgray}{89.4/135} \\
1 goal, 60\% & 100.0/31 & 100.0/48 & 100.0/31 & \cellcolor{bestgreen}\textbf{100.0/29} & 100.0/31 & 100.0/32 & 100.0/30 & 100.0/46 \\
3 goals, 20\% & \cellcolor{bestgreen}\textbf{100.0/136} & 100.0/410 & 100.0/212 & 100.0/411 & 100.0/379 & 100.0/187.5 & 100.0/165.5 & 100.0/436 \\
3 goals, 40\% & \cellcolor{bestgreen}\textbf{100.0/159} & 100.0/417.5 & 100.0/203 & 100.0/209.5 & 100.0/220 & 100.0/186.5 & 100.0/186 & 100.0/581 \\
3 goals, 60\% & 100.0/83 & 100.0/138 & 100.0/80 & \cellcolor{bestgreen}\textbf{100.0/72} & 100.0/77 & 100.0/83 & 100.0/83 & 100.0/137.5 \\
\midrule
\multicolumn{9}{@{}l}{\textit{Static mazes: learning from repeated experience (10 cycles)}} \\
1 goal, 0\% & \cellcolor{bestgreen}\textbf{100.0/38} & 97.7/202 & 99.9/116 & \textcolor{lowgray}{0.0/--} & \textcolor{lowgray}{0.0/--} & 100.0/90 & 100.0/90 & 100.0/50 \\
1 goal, 20\% & \cellcolor{bestgreen}\textbf{99.7/38} & 91.4/262 & 99.6/126 & \textcolor{lowgray}{6.9/30} & \textcolor{lowgray}{32.2/24} & 100.0/92 & 100.0/86 & 100.0/52 \\
1 goal, 40\% & \cellcolor{bestgreen}\textbf{97.5/36} & \textcolor{lowgray}{66.2/389} & 96.6/180 & \textcolor{lowgray}{49.0/46} & \textcolor{lowgray}{48.7/42} & 100.0/84 & 100.0/82 & 100.0/54 \\
1 goal, 60\% & \cellcolor{bestgreen}\textbf{99.6/22} & \textcolor{lowgray}{77.8/340} & 100.0/82 & 94.9/34 & \textcolor{lowgray}{56.5/34} & 100.0/38 & 100.0/36 & 100.0/40 \\
\bottomrule
\end{tabular*}
\par\smallskip
\begin{minipage}{1\textwidth}\footnotesize
Each cell of the table shows the fraction of completed routes (\%) / median length of the completed routes (steps). Among methods completing at least 95\% of routes, the method with the shortest median route in each row is shown in bold on a green background; results with fewer than 90\% of routes completed are shown in gray.  Static mazes: the astrocyte-induced short-term memory capacity is set to 100 steps; Q-learning is trained from scratch within each trial. Dynamic mazes: at each step, 20\% of all walls are moved to other locations, while the number of walls is kept constant and all goals remain reachable (Section~\ref{sec_dyn_cmp}). Static mazes, learning from repeated experience: a fixed start and goal, with a new maze in each session; SNAN and Q-learning carry their synaptic weights or Q-table over to the next cycle, whereas the other methods do not reuse previous experience (Section~\ref{sec_learn_cmp}). Memory requirements within a trial: variables retained between consecutive steps of a trial in addition to the current position; $n=N^2$ positions and $E=2N(N-1)$ transitions between neighboring positions of an open $N\times N$ grid. RW: random walk; RM: random mouse; WF: wall follower; Q: tabular Q-learning.
\end{minipage}
\end{table*}

\subsubsection{Memory requirements}

Since the aim of this work is to develop a navigation system based on a bioinspired neuron--astrocyte network for subsequent hardware implementation on neuromorphic computing devices, and to compare it with classical maze-solving methods, the memory requirements of each method must also be considered. Therefore, in addition to the completion rate and route length, we assessed what each strategy must store between consecutive steps of a trial beyond the current position and how many memory cells are required for this in an open $10\times10$ grid. For an open $N\times N$ grid, let $n=N^2$ denote the number of positions and $E=2N(N-1)$ denote the number of transitions between neighboring positions; these quantities are used to quantify the memory requirements within a trial. The results of this assessment for all compared methods, together with a description of the stored information, are given in the first row of Table~3. SNAN stores one analog variable per agent position---astrocytic activity (intracellular Ca$^{2+}$ concentration)---which represents astrocyte-induced short-term memory. Synaptic weights are not included in this count, because within a single trial the performance of SNAN with astrocytes did not depend on synaptic plasticity (Section~\ref{sec_results1}), and the membrane potentials are reset in each decision window. Among the strategies that store information about the visited environment, SNAN requires the fewest memory cells: 100 versus 282 for Tr\'emaux's algorithm, 360 for Q-learning, and 460 for Tarry's algorithm. SNAN stores one variable per position, whereas the other strategies store variables per transition or per state--action pair, so the advantage of SNAN grows with the number of neighbors of each position. Only strategies with constant memory require fewer cells, but they pay for this with long routes and/or low reliability. In addition, astrocytic memory is local, because each astrocyte is associated with the position it encodes, and self-erasing, because it decays after astrocytic modulation ends, without any explicit reset.

\subsubsection{Static mazes}

The \textit{Static mazes} block of Table~\ref{tab_compare} shows the results in static environments without learning (1000 independent trials per row). For SNAN, the astrocyte-induced short-term memory capacity was fixed at 100 steps. SNAN combines short routes with high reliability. It completed at least 96.2\% of routes in all conditions. In the single-goal task in open and sparsely obstructed grids, and in the multi-goal task in the open grid, it produced the shortest routes of all reliable strategies, i.e., those completing at least 95\% of routes: 60 and 68 steps at 0\% and 20\% wall density versus 75 and 79 steps for Tarry's algorithm, and 114.5 versus 154 steps in the three-goal task. In open space, the depth-first exploration strategy of the classical algorithms (Tr\'emaux’s and Tarry’s) leads to long detours, whereas the astrocyte-induced short-term memory of recently visited cells drives the agent toward unexplored regions without a fixed order of traversal. In open and sparse grids, strategies that do not store information about the environment are either slow or unreliable. Random-walk routes were 2.4--3.3 times longer than those of SNAN, and the wall follower and the Pledge algorithm completed at most 40.3\% of routes. Q-learning, trained from scratch for each task, required 1.8--2.3 times more steps than SNAN. In denser mazes and in most multi-goal tasks, Tr\'emaux's and Tarry's algorithms produced shorter routes than SNAN, with the largest difference observed in the multi-goal task at 40\% wall density. Their advantage rests on an exact, permanent record of every traversed passage, which enables systematic traversal of the maze. SNAN keeps only a decaying trace of recently visited positions. With a fixed memory duration of 100 steps, this trace can also suppress corridors that must be traversed again to reach the remaining goals (Section~\ref{sec_results4}).

\subsubsection{Dynamic mazes}\label{sec_dyn_cmp}

\begin{figure}
\centering
\includegraphics[width=1\columnwidth]{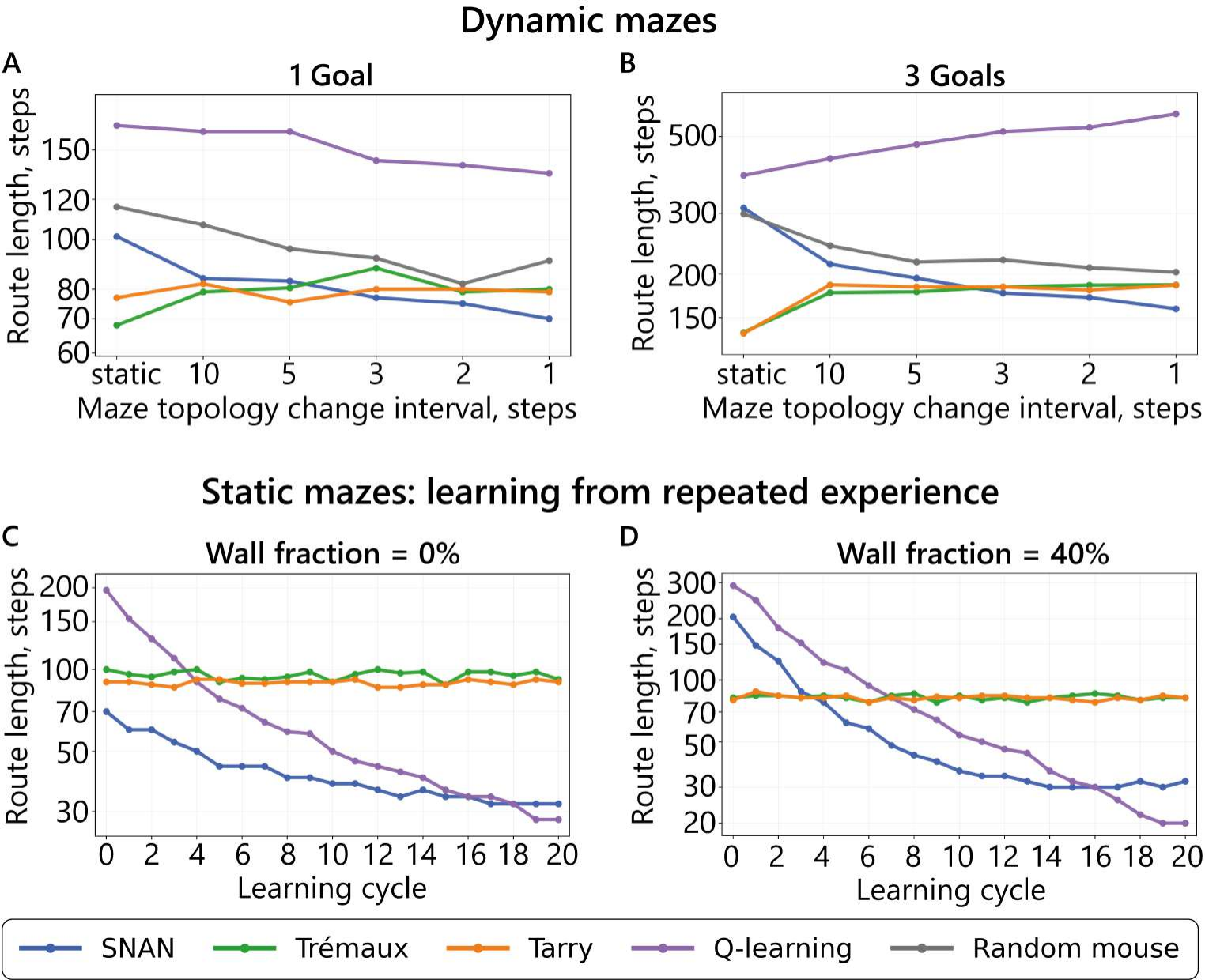}
\caption{Comparison of the neuromorphic navigation system with classical navigation strategies. (A, B) Performance of neuromorphic navigation system in dynamic environment. Navigation in dynamic mazes with 40$\%$ wall fraction during one trial: median route length as a function of the maze topology change interval for the task with one goal (A) and three goals (B). (C, D) Learning from repeated experience in the open grid (C) and in static
random mazes with a wall fraction of 40$\%$ (D). The median route length is shown after each of the 20 learning cycles. SNAN (with astrocyte-induced
short-term memory and memristive STDP) and Q-learning carry their learned parameters (synaptic weights or Q-tables) over to the next cycle, whereas
Tr\'emaux's and Tarry's algorithms do not use previous experience.  (10×10 grid, 1000 trials per task identical for all methods) }
\label{fig8}
\end{figure}

The neuromorphic navigation system proposed in this study combines astrocyte-induced short-term memory and STDP-induced long-term memory. We hypothesized that our system would be advantageous in dynamic environments where the maze topology changes during the agent's navigation, whereas the permanent marks of classical algorithms (e.g., Tr\'emaux's and Tarry's) become obsolete. To test this, 20\% of the walls (in all cases considered below) were relocated every few steps during the trial; we refer to the number of steps between changes as the maze-topology change interval. The number of walls was kept constant; all remaining goals remained reachable; and the sequence of changes was the same for all methods. If a classical agent attempted to use a passage that had just been closed, it made a random admissible move instead. We compared the performance of our neuromorphic agent with that of four previously considered navigation methods---Tr\'emaux, Tarry, random mouse, and Q-learning---on a dynamic maze-navigation task with a wall fraction of 40\%. In this task, walls were relocated at intervals of 10, 5, 3, 2, or 1 step (Fig.~\ref{fig8}A, B). The agent was allowed up to 1000 steps per trial in the one-goal task and up to 5000 steps per trial in the multi-goal task. As the changes became more frequent, SNAN became more efficient, whereas Tr\'emaux's and Tarry's algorithms lost their advantage. In the single-goal task, the median route length of SNAN decreased from 101.5 to 70 steps, whereas those of Tr\'emaux's and Tarry's algorithms increased slightly (from 68 and 77 to 80 and 79 steps, respectively; Fig.~\ref{fig8}A). In the multi-goal task, the median route length of SNAN decreased from 311 to 159 steps when the walls changed at every step, while those of Tr\'emaux's and Tarry's algorithms increased from 136 and 135 to 186.5 and 186 steps, respectively (Fig.~\ref{fig8}B). The memoryless random mouse also benefited from the changes, but to a lesser extent than SNAN did (from 116 to 91 steps for one goal and from 299 to 203 steps for three goals), so the improvement in SNAN was not explained by the changing maze alone. The decaying astrocyte-induced short-term memory prevents cycling over the recently traversed part of the route but does not retain outdated information about passages that no longer exist. We also compared all the maze-navigation methods considered above in dynamic mazes with wall fractions of 20\%, 40\%, and 60\% and a topology-change interval of 1 step (Table~\ref{tab_compare}). In mazes with wall fractions of 20\% and 40\%, SNAN produced the shortest median route lengths among all methods in both tasks and completed 100\% of routes. The neuromorphic agent's routes were significantly shorter than those of all other methods ($p<0.01$), except in the case of Tarry's algorithm in the single-goal task at a 40\% wall fraction ($p=0.06$). In mazes with high obstacle density (60\% wall fraction), where only a few alternative paths exist, the wall-follower method had a slight advantage over our system. Specifically, the median route length of SNAN was two steps longer than that of the wall follower in the one-goal task (31 versus 29 steps) and 15\% longer in the three-goal task (83 versus 72 steps).

\subsubsection{Learning from repeated experience}\label{sec_learn_cmp}

Next, we compare the performance of the proposed neuromorphic navigation system with the Q-learning algorithm in the maze route-learning task with
repeated experience, analogous to that considered in Section~\ref{sec_results6}. The experimental session consisted of 20 consecutive cycles of agent
navigation in the same grid environment (an open grid and mazes with wall fractions of 0--60$\%$), without reinitializing the network's synaptic weights
at the beginning of each cycle. To assess the dynamics of the agent's navigation performance as it learned the route over the course of the session, we tested the agent after each of the 20 cycles using the synaptic weight values from the preceding cycle, with astrocyte-induced short-term memory enabled. For each grid environment and each agent, we conducted 1000 independent experimental sessions with random initial synaptic weights and a fixed start--goal pair. Performance was evaluated using the route length and the fraction of completed routes---i.e., the proportion of trials in which the agent reached the goal within the maximum allowed number of steps (1000). A comparison of the performance of all agents in this task is presented in the last block of Table~\ref{tab_compare}. After 10 learning cycles, SNAN produced the shortest median route length compared with all reliable navigation strategies in all maze topologies under consideration (22--38 steps versus 40--54 steps for Q-learning and 36--92 steps for Tr\'emaux's and Tarry's algorithms) and completed at least
97.5$\%$ of routes (Table~\ref{tab_compare}). In pairwise comparisons, the neuromorphic agent produced a shorter route than Q-learning in 55--72$\%$ of
tasks and than Tr\'emaux's and Tarry's algorithms in 68--77$\%$ of tasks ($p<0.01$). Figure~\ref{fig8} shows a comparison of the rate of decrease in
the median route length as a function of the learning cycle number for the neuromorphic navigation system, Q-learning, and Tr\'emaux's and Tarry's
algorithms in an open grid and in mazes with a 40$\%$ wall fraction. SNAN learned faster than Q-learning, which started from much longer routes (196 and
290 steps), and its median route remained shorter up to the 15th cycle. With longer training, however, Q-learning converged to shorter routes (28 versus 32
steps in the open grid and 20 versus 32 steps in the mazes), as expected for a method that optimizes an explicit reward. Clearly, Tr\'emaux's and Tarry's
algorithms did not improve with repeated experience.

\section{Hardware implementation}\label{sec_hardware}

\begin{figure}
\centering
\includegraphics[width=1\columnwidth]{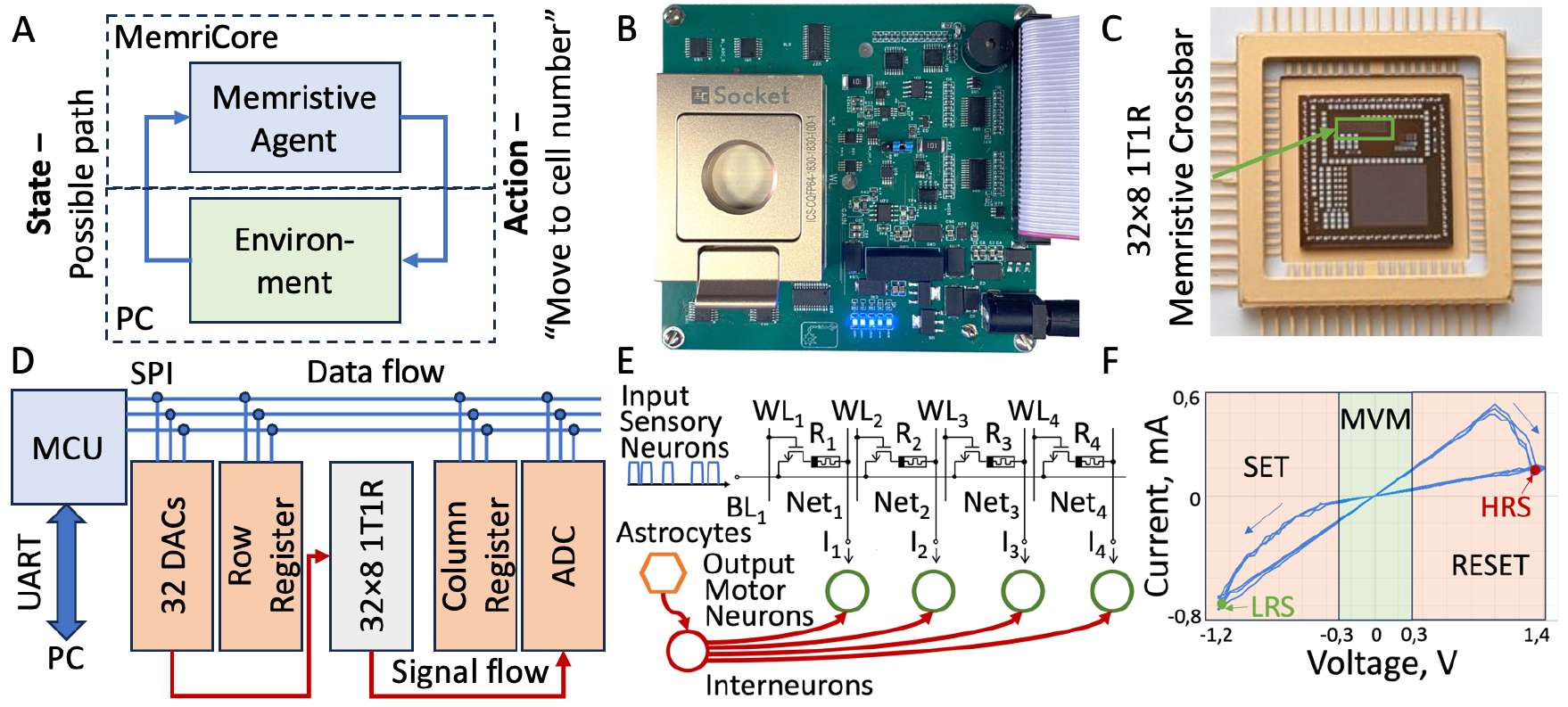}
\caption{Hardware implementation of the agent. (A) Experimental setup diagram. Some of the agent's functions are performed on the PC, and the other part in the memristor-chip-based platform. (B) Photo of the memristor-chip-based MemriCore platform. (C) A photograph of the memristor chip. Green indicates a 32×8 1T1R crossbar array. (D) Block diagram of the memristor-chip-based platform. (E) Memristors in the crossbar-array are the synapses of the output motor neurons. (F) IV-curves of memristors from the crossbar array.}
\label{fig9}
\end{figure}

To demonstrate in principle the operation of the agent in neuromorphic hardware, an experimental setup was created, consisting of a personal computer (PC) and a MemriCore board based on a memristor chip (Fig.~\ref{fig9}A) \citep{shchanikov2023modeling}. Some key functions of the agent, namely synapses and motor neurons, were implemented on the MemriCore board using a memristor crossbar array and the MemriBoard framework \citep{shchanikov_memriboard}, while the rest of the agent's functionality and the environment were handled by the PC. Data exchange between the MemriCore and the PC was carried out via a UART interface through an MIK 32 Amur (Risc-V) microcontroller (MCU), which also manages the operation of all components. The MemriCore board has a socket (Fig.~\ref{fig9}B) for installing a memristor chip in a 64-pin metal-ceramic package.

For the experiments, a chip with a 32x8 cell memristor crossbar array (Fig.~\ref{fig9}C) was used, where each cell includes an n-channel CMOS transistor and a metal-oxide memristor based on yttria-stabilized zirconia \citep{mikhaylov2020multilayer}. The use of a memristor other than the one used in the simulation part of the work is due to the need to verify diversification of the hardware implementation of SNAN. Such diversification, if successful, proves that only the causal nature of local STDP-like plasticity is crucial, and not the specific type of its implementation. The MemriCore board features an SPI bus (Fig.~\ref{fig9}D) connected to four devices: a 32-channel 12-bit DAC, a single-channel 14-bit ADC, and two registers for row and column addressing. The registers control the multiplexers and allow operation with either a single cell or an entire column, enabling the scalar multiplication of two 32-element vectors.

In a neural network, 4 sensory neurons at the corner cells of the field have connections with 2 motor neurons, 16 more along the borders with 3, and 16 in the center with 4; thus the total number of synapses is 120. Since spikes from one sensory neuron are the same for all synapses connecting this neuron to the motor ones, the memristors of such synapses can be arranged in the columns of the crossbar array, as shown in Fig.~\ref{fig9}E, and the connection between the sensory neuron and 2, 3 or 4 motor ones will be set by applying 3.3 V to the corresponding WL. In this case, the input signal to the DAC needs to be set once for a single spike, and the ADC must be polled sequentially after writing the column bit to the column register ($WL_1$, $WL_2$, ...) for all motor neurons. During inference, the amplitude of the spikes must not exceed 0.3 V, as shown in Fig.~\ref{fig9}F, to avoid inadvertently changing the memristor resistance. For writing weights, a range from 0.3 V to 1.2--1.4 V can be used.

To conduct the experiments, the agent was trained 10 times for a 6×6 field to find the path from cell 18 to cell 0 with random initial weights each time. The resulting sets of weights were then mapped to the memristive crossbar array as follows: from a sparse 36×36 matrix (in the case of fully connected grid cell matrix), non-zero values for 120 weights were extracted and transformed into a 15×8 conductance matrix (to place them in crossbar array 32×8). For each weight, an equivalent target conductance was calculated, and these values were programmed into the crossbar array.

\begin{figure}
\centering
\includegraphics[width=1\columnwidth]{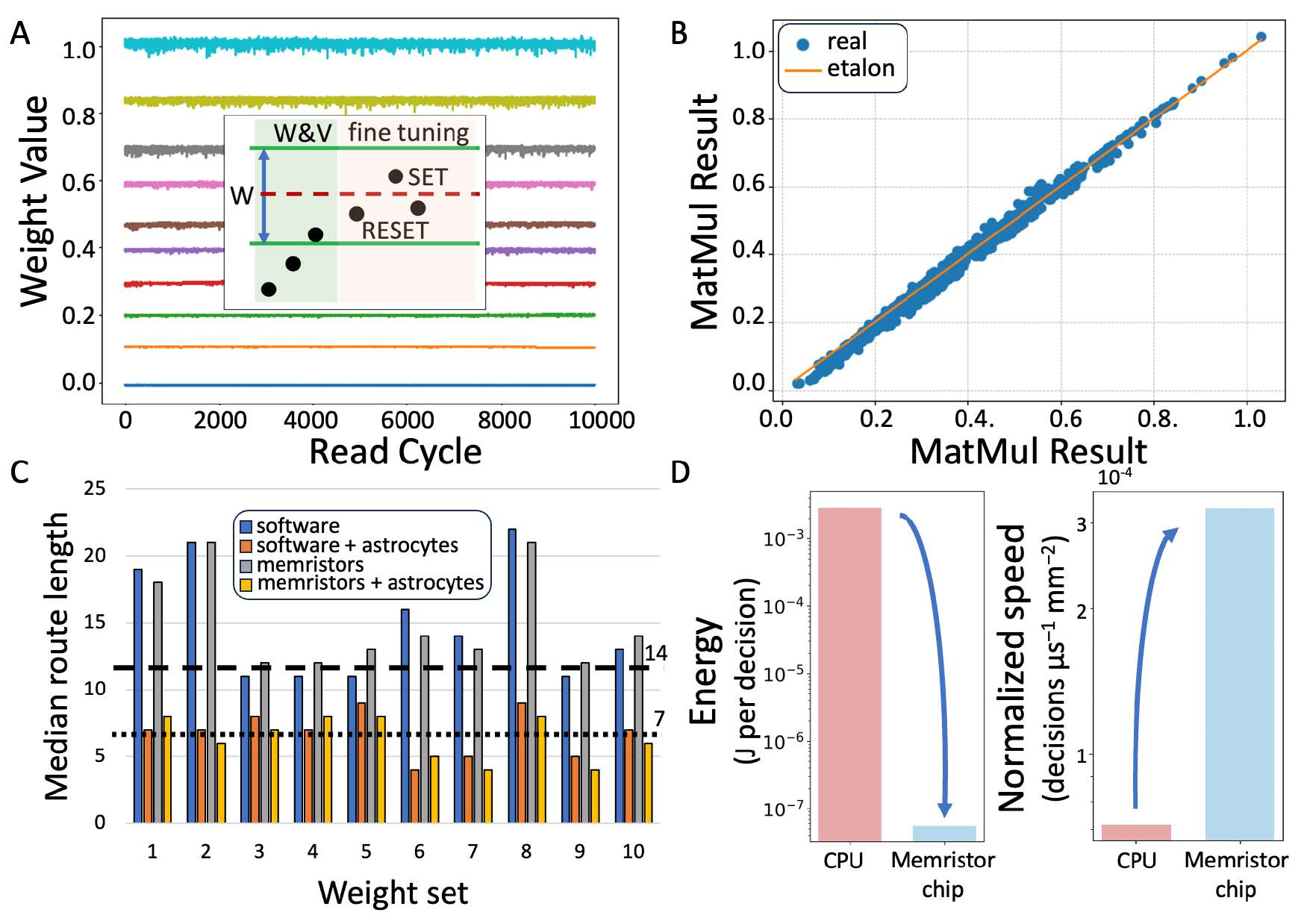}
\caption{Results of experiments with hardware. (A) An example of writing weights in the range from 0 to 1 in a memristive crossbar array and holding them for 10,000 read cycles; no relaxation is observed. The box shows a two-step weight programming algorithm - write and verify (W\&V) to get into the weight window and fine tuning to get closer to the center. (B) The error of the matrix multiplication in the MemriBoard (blue dots) relative to the reference result (orange line). (C) The result of the agent's movement on a 6×6 field for a software model and a hardware implementation on memristors with and without astrocyte modulation (the dashed line shows the median path length for each set of weights without astrocytes and the dotted line - for the experiments with the astrocytic modulation). (D) Estimates of power consumption and performance for a CPU software and hardware implementation on memristors.}
\label{fig10}
\end{figure}

The weight programming was carried out in two stages, differing only in the stopping criterion. The programming signal consisted of a sequence of ``write'' and ``read'' pulses (write and verify method). The amplitude of the write pulses was increased from 0.3V to 1.4V, first in RESET then in SET, while the amplitude of the ``read'' pulse was 0.3V. The result of the ``read'' operation was checked against the signal's stopping condition. In the first stage, the condition was $G_t - 10\% \le G_c \le G_t + 10\%$, where $G_t$ is the target memristor conductance and $G_c$ is the current measured conductance. In the second stage, there were two stopping conditions: first, RESET pulses were applied until $G_c \le G_t$, followed by SET pulses until $G_c \ge G_t$. This method allows the memristor's current conductance to be brought as close as possible to the target, as shown in Fig.~\ref{fig10}A.

Subsequently, the agent was launched 10 times for each weight set in order to test repeatability when performing analog calculations. Since the multiplication in the memristive crossbar array is performed in an analog manner, this introduces errors compared to the software implementation, an example of which is shown in Fig.~\ref{fig10}B. Errors arise both during weight programming and when applying spikes, which inevitably contain noise. Conducting 10 experiments for 10 different sets of weights makes it possible to assess the reproducibility and repeatability of the results. The experimental results are shown in Fig.~\ref{fig10}C for four different agent launch scenarios: software implementation, software implementation with astrocytic modulation, memristor-based implementation, and memristor-based implementation with astrocytic modulation. The figure is divided into 10 bar sets for each set of weights and for each launch scenario. The solid lines represent the median path length of the agent. Two conclusions can be drawn from the experimental results. First, despite the inaccuracies of analog computation, the hardware implementation keeps the median path length relative to the software implementation. The second conclusion is that, in both the software and hardware implementations, astrocytic modulation significantly shortens the agent's path to the target point (7 steps with and 14 steps without astrocytic modulation).

The advantages of memristors in hardware implementation of neuromorphic systems include good energy efficiency while maintaining high performance. Comparison results show that although memristor-based neuromorphic systems still exist as laboratory prototypes and are inferior to industrially manufactured processors in terms of the number of synapses and absolute performance, in relative terms they can provide orders of magnitude better energy efficiency and performance \citep{mikhaylov2023neuromorphic}. In this work, we also evaluated these metrics in comparison with our full software implementation on a PC with an Intel Core i5 12450H CPU according to the methods described by \citep{liu2025memristor}. 

According to \citep{intelbenchmark}, the Intel Core i5-12450H CPU has a computing power of 40.3 GOPS, an average frequency of 3.2 GHz, a typical TDP of 45 W, and its chip area is 217 mm$^{2}$. This means that one agent step for a field size of (6,6), if the simulation time step is set to 1 ms, would be executed in (36*36*2*1000) / 40.3 GOPS = 64.4 $\mathrm{\mu}$s. Normalizing this value by the chip area yields 1 / 64.4 $\mathrm{\mu}$s / 217 mm$^{2}$ = 7.16e-5 decisions $\mathrm{\mu}$s$^{-1}$ mm$^{-2}$. The MemriCore board is equipped with AD5328 DACs with a settling time of 6 $\mathrm{\mu}$s, TLC3541 ADCs with a 200 kSPS (kilosamples per second) sampling rate, and 74HC595PW column registers. For a (6,6) field, 120 output motor neurons are needed, which fits within a (15,8) subsection of the crossbar array. Firing one spike onto 15 rows and reading the 8 columns for each would take 700 $\mathrm{\mu}$s. The average number of spikes per step is 12, and the area of the 32x8 1T1R crossbar array is 0.368 mm$^{2}$, so the normalized speed would be 700 $\mathrm{\mu}$s * 12 / 0.368 mm$^{2}$ = 3.23e-4 decisions $\mathrm{\mu}$s$^{-1}$ mm$^{-2}$, which is 4.5 times higher than on the CPU (Fig.~\ref{fig10}D). When transitioning to an integrated version of the memristive crossbar-array with DACs and ADCs, the clock frequencies can be increased to hundreds of MHz, which will make its speed even greater, up to three orders of magnitude higher than that of the CPU.

Energy consumption is calculated as the execution time of one agent step multiplied by the TDP. For the Intel Core i5-12450H CPU, this is 64.4 $\mathrm{\mu}$s * 45 W = 0.0029 J. For the memristor crossbar array, we estimate the energy consumption using the formula for the work of the electric field, with an average motor neuron synapse resistance of 1.5 k$\Omega$ and a spike voltage of 0.1 V. This yields 0.1 * 0.1 / 1500 * 0.0084 = 56 nJ per step, which is 51,739 times more energy-efficient than the CPU (Fig.~\ref{fig10}D). When comparing the computing units, we do not consider the peripheral equipment present for both the CPU (motherboard, RAM, power supplies) and the memristor crossbar array (DACs, ADCs, MUXs, MCU), considering the speed and consumption only for performing MAC operations on the data, not accounting for how this data arrived at the computing unit. Nevertheless, this first in-principle estimation shows that realizing SNAN in dedicated neuromorphic hardware with locally executed MAC calculations can be significantly more efficient than in the case of general-purpose out-of-memory computing hardware.

\section{Discussion}\label{sec_discussion}
\subsection{Exploration in reinforcement learning and neuromorphic navigation} \label{sec_discussion1}

The exploration-exploitation dilemma represents one of the most fundamental challenges in reinforcement learning and sequential decision-making, lying at the heart of an agent's ability to adapt to unknown environments. Striking the right balance between gathering information about the world and exploiting current knowledge to maximize reward is essential for any intelligent system operating in real-world conditions, from autonomous robots navigating unfamiliar terrain to animals foraging in the wild. Despite decades of research, achieving this trade-off efficiently and robustly remains an open problem, particularly in resource-constrained settings where computational efficiency is paramount.

Classical approaches to exploration laid the essential groundwork for understanding this balance. Methods such as $\varepsilon$-greedy \citep{Watkins1989, Sutton1998}, softmax (Boltzmann exploration) \citep{Luce1959, Sutton1998}, Upper Confidence Bound (UCB) \citep{Lai1985, Auer2002}, and Thompson sampling \citep{Thompson1933} have been widely adopted owing to their simplicity and theoretical grounding. However, each has inherent limitations that restrict its applicability in complex environments. $\varepsilon$-greedy performs undirected random exploration irrespective of informational value, leading to inefficient coverage of large state spaces. Softmax enables probabilistic action selection weighted by current value estimates, but its performance critically depends on the temperature parameter $\tau$, with poor tuning leading either to random exploration or to premature convergence. UCB and Thompson sampling, while more sophisticated, require maintaining global statistics and scale poorly to multi-step tasks with large state spaces \citep{Auer2002, Slivkins2019}.

Modern approaches have substantially advanced exploration capabilities in high-dimensional environments, but this progress has come at a significant computational cost. Intrinsic motivation methods, including count-based exploration with pseudo-counts \citep{Bellemare2016, Ostrovski2017} and prediction-based approaches such as ICM \citep{Pathak2017} and RND \citep{Burda2018}, enable effective exploration in complex domains such as Atari games. However, as \citep{Ostrovski2017} note, pseudo-count methods require expensive density model updates at every step. Prediction-based methods similarly require training additional networks and computing  prediction errors continuously, imposing a substantial computational burden \citep{Pechac2024}. Hierarchical methods employing temporal abstractions \citep{Sutton1999, Bacon2017} require multi-level policy optimization and additional mechanisms for discovering abstractions. Meta-reinforcement learning algorithms \citep{Duan2016, Zintgraf2020} require training recurrent architectures on large numbers of tasks. More fundamentally, \citep{Norman2024} identified a critical failure mode even for state-of-the-art meta-RL methods: they are unable to learn to explore in situations where optimal behavior requires forgoing immediate reward for greater future gains. Their proposed solution — learning two separate policies, an explore policy and an exploit policy — partially resolves this issue, but at the cost of two independent learning processes and an external switching mechanism, which further increases system complexity. Moreover, like their classical counterparts, most of these approaches assume stationary reward distributions and struggle in non-stationary environments where optimal policies shift over time \citep{Durstewitz2025}. Thus, modern methods achieve impressive results, but through computationally intensive architectures that are ill-suited for energy-constrained edge devices and autonomous robots operating in dynamic, unpredictable environments.

Amidst these challenges, neuromorphic computing offers a fundamentally different paradigm inspired by the brain's exceptional energy efficiency \citep{Chowdhury2025, Novo2024, Durstewitz2025}. Spiking Neural Networks (SNNs) encode information in sparse, asynchronous spikes, co-locating computation and memory \citep{Maass1997, Roy2019}. Dedicated platforms such as Intel Loihi \citep{Davies2018}, IBM TrueNorth \citep{Akopyan2015}, and SpiNNaker \citep{Furber2014} achieve order-of-magnitude power reductions over traditional GPUs. In vision-based navigation, event cameras paired with SNNs enable real-time optical flow \citep{Benosman2014, Barranco2015}, ego-motion tracking \citep{Rebecq2017, Zhou2021}, and depth estimation \citep{Rancon2022}. Architectures such as Spike-FlowNet \citep{Lee2020} — a hybrid SNN-ANN — outperform corresponding ANN-based methods in optical flow prediction accuracy while providing significant computational efficiency. Fully spiking architectures such as Adaptive-SpikeNet \citep{Kosta2023} go further, achieving 20$\%$ lower average endpoint error than ANN-based models of comparable size, with substantial savings in parameters (48×) and energy (10×). For higher-level navigation, RatSLAM \citep{Ball2013, Milford2008} and NeuroSLAM \citep{Yu2019} employ continuous attractor networks inspired by hippocampal place and grid cells, achieving robust SLAM without GPS. Models incorporating place cells and grid cells support goal-directed navigation \citep{Banino2018, Edvardsen2017, Chen2021}. However, all these systems rely on rich sensory input (visual flow, depth, odometry) and often on an explicit goal representation. The question of how an agent might explore and learn when no such information is available — when it has no map and no goal signal and knows only the identity of its current position—remains largely unaddressed in spiking and neuromorphic navigation models. In this regime, an agent faces the exploration problem in its most extreme form: it must rely on internal dynamics to avoid cyclic behavior and discover the structure of its environment. While SNNs naturally incorporate temporal dynamics through membrane leak, these dynamics are far too fast to bridge tens or hundreds of decisions; what is needed is a slower short-term adaptive memory that can temporarily suppress recently visited states, thereby enabling the agent to break out of cycles and continue exploring. Astrocytic modulation has recently emerged as a biologically plausible candidate for such a mechanism, with studies demonstrating its role in short-term plasticity for working memory \citep{Gordleeva2021, Gordleeva2025} and in sensor-based navigation \citep{Liu2019}.

\subsection{Dual-timescale memory and Topological-Context Memory} \label{sec_discussion2}

Building on these foundations, the proposed SNAN resolves the exploration-exploitation dilemma in this minimal-observation regime. Our approach combines two local plasticity mechanisms operating on different timescales: STDP provides long-term memory for exploiting accumulated experience, while astrocyte-mediated suppression creates a short-term ``I  have been here recently'' effect that temporarily suppresses moves into recently visited cells. Within a single trial, the performance gain is due mainly to this astrocytic trace (Section ~\ref{sec_results1}): it raised goal completion from 45.9\% to 99.9\% in the open 10×10 grid and from 13.0-–45.9\% to 96.2--100\% in static mazes, and on task instances solved by both agent variants it shortened median routes 5.8--15-fold. The duration of astrocytic activity serves as a single interpretable parameter controlling the exploration-exploitation balance. With very short memory (1--10 steps), the agent frequently becomes trapped in cyclic behavior (75.5--92.9$\%$ of completed routes in the open 10×10 grid); from 30 steps upward, the success rate is at least 98.9$\%$, and in the multi-goal task with return, durations of 50--500 steps yield 96.9--100$\%$ of completed routes at all wall fractions. Further increases preserve the success rate but lengthen the routes, because the agent then also suppresses corridors that must be traversed again; the optimal duration (about 40–200 steps in our tasks) depends on the size and wall density of the environment (Section~\ref{sec_results4}). Nevertheless, the success rate remains high over a broad range of durations, so that a single parameter with a wide operating window replaces the fine-tuning of multiple sensitive coefficients. This window is, however, environment-dependent: the calcium time constant was increased from 2 s to 22 s when moving from the 3×3 to the 10×10 grid, and a mechanism that adapts the memory duration to the environment remains to be developed. 

We refer to this form of working memory as Topological-Context Memory. In conventional approaches, context is provided to the agent as a global history or a latent state vector that is processed in full at every inference step (Section~\ref{sec_Intro}). In SNAN, by contrast, the context is distributed across astrocytes, each coupled to the neurons that encode one position, so that at each decision only the astrocytes associated with the candidate next positions affect the computation. Immediate sensory information is thus processed together with contextual data encoded in topologically specific astrocyte-neuron interactions, while the other astrocytes keep their traces in a background mode with respect to the current inference step. The global working memory is then the union of these local traces. In the grid tasks studied here, network topology coincides with spatial adjacency, but the principle is not restricted to space. For example, an agent that solves several tasks and periodically switches between them could keep a short-term context for each task in such a background mode and use only the context of the current task, together with sensory data, at the current inference step. Such an organization shortens the context involved in immediate computations. Whether this also reduces the overall computational cost remains to be demonstrated: in our simulations, astrocytic variables were integrated with the same time step as the neurons, and savings would require event-driven or slower updating of the astrocytic layer, which its longer timescale permits in principle.

\subsection{Relation to classical navigation strategies and other exploration mechanisms} \label{sec_discussion3}

The comparison with classical navigation strategies (Section~\ref{sec_results7}) illustrates the practical consequences of this design. Exact graph-exploration algorithms, such as Tr\'emaux's and Tarry's, keep a permanent record of every traversed passage; they are therefore guaranteed to reach every reachable goal and are hard to beat in static mazes of intermediate and high density, and in most multi-goal tasks, when no previous experience is available. The astrocytic memory of SNAN is smaller (one variable per position rather than per transition, i.e., 2.8--4.6 times fewer stored variables in the open 10×10 grid), local and self-erasing. It remains useful when the environment changes, whereas permanent passage marks become obsolete: as walls were relocated more frequently, the median routes of SNAN became shorter (from 101.5 to 70 steps with one goal and from 311 to 159 steps with three goals), whereas those of Tr\'emaux's and Tarry's algorithms became longer. It is also complemented by reward-free synaptic learning: after ten learning cycles, the median routes of SNAN were 1.6--2.4 times shorter than those of Tr\'emaux's and Tarry's algorithms and shorter than those of Q-learning at all tested wall fractions. With longer training, Q-learning, which optimizes an explicit reward, converged to shorter routes (Section~\ref{sec_learn_cmp}), so the advantage of SNAN lies in fast improvement from little experience rather than in asymptotic optimality. Both advantages are most pronounced in open and sparsely obstructed environments, where many alternative paths exist; in dense mazes with few alternative paths, even simple strategies such as wall following remain competitive. These properties are particularly relevant for autonomous agents operating in non-stationary environments and for neuromorphic hardware, in which memory has to be co-located with computation.

Functionally, the astrocytic trace is related to several known mechanisms that discourage revisits and have been developed in different domains. In visual search, inhibition of return biases attention and gaze away from recently inspected locations for several seconds, favoring novel ones \citep{Klein2000}. In combinatorial optimization, tabu search forbids moves that would reverse recent moves for a fixed number of iterations, which prevents the search from cycling back to recently visited solutions \citep{Glover1989}. Closest to our setting, in multi-robot coverage of an area represented as a graph, robots leave evaporating traces and move toward the neighbor with the weakest trace, that is, the least recently visited one \citep{Wagner1999}. In SNAN, the same principle is implemented by astrocyte-induced short-term memory, whose duration is set by the calcium dynamics rather than by an explicit counter or a comparison of trace strengths.

Using the VTEAM model \citep{Kvatinsky2015} parameterized with experimental memristor characteristics supports physical realizability without additional tuning of the learning rule \citep{Kulagin2025, Rybka2024}. Unlike prior work on three-factor STDP rules \citep{Fremaux2016, Tihomirov2025, Vlasov2023} and memristive implementations \citep{Kulagin2025, Rybka2024}, which assume full observability, SNAN demonstrates effective exploration in the case of partial observability of state. Conceptually, it implements a temporal analogue of the separation of explore and exploit policies in First-Explore \citep{Norman2024}: STDP serves as the long-term ``exploit'' policy, while astrocytic suppression provides a short-term ``explore'' policy that temporarily blocks familiar actions in favor of unvisited ones. This switching is governed by intrinsic calcium dynamics rather than an external mechanism, integrating both policies within a single architecture acting on a common set of motor neurons. This integration simplifies learning and facilitates hardware implementation.

\subsection{Towards a full neuromorphic hardware implementation} \label{sec_discussion4}

Implementing the full SNAN on a memristive crossbar would require addressing several challenges. First, the astrocytic calcium dynamics constitute a slow analog process with time constants of hundreds of milliseconds to seconds, fundamentally different from the fast, event-driven operation of memristive synapses; realizing such slow temporal integration would demand additional analog circuit elements or a hybrid approach in which calcium variables are emulated digitally and only their modulatory effect is applied to the crossbar. Second, the astrocyte-to-interneuron interaction involves a thresholded, state-dependent modulation of neuronal excitability rather than a synaptic weight change, and translating this into a memristive element would require a device whose conductance is controlled by a separate, slow-varying terminal, which remains an emerging technology. Third, the calcium threshold and the resulting suppression must be maintained locally at each unit, implying a per-cell analog memory element in addition to the synaptic array. Nevertheless, recent demonstrations of volatile memristors with tunable relaxation times and transistor-integrated memristive arrays suggest that astrocyte-inspired slow dynamics could be realized in hardware, either through dedicated volatile devices or through a hybrid CMOS–memristor architecture. We therefore regard the full memristive implementation of the SNAN, including astrocytic modulation, as a promising but non-trivial direction for future work.

\subsection{Limitations and outlook} \label{sec_discussion5}

Other limitations suggest further directions. First, the state space is discrete and each position is represented by a dedicated network unit, so the network size grows linearly with the number of states; distributed representations such as grid cells \citep{Banino2018}, continuous attractor networks \citep{Samsonovich1997}, or abstract state representations could be integrated. Second, astrocytic suppression is reactive---it prevents revisits but does not actively guide exploration toward informative regions, motivating hybridization with predictive models. Third, perfect state identification is assumed; the effect of noisy or ambiguous observations on the astrocytic trace requires systematic investigation. Fourth, the evaluation was limited to grids of up to 10×10 cells and to baselines that start without prior knowledge; tests in larger and continuous environments and comparisons with recurrent and meta-RL agents after pre-training are needed to delineate the scope of the approach. Fifth, hierarchical organization \citep{Sutton1999, Dzhivelikian2022} could integrate astrocytic mechanisms for multi-timescale exploration, particularly given evidence that distinct neural circuits regulate directed versus random exploration \citep{Asch2025}.

In a broader context, the proposed approach shows that the exploration-exploitation dilemma can be addressed not only at the level of algorithmic augmentations to standard RL architectures, but also through the embedding of specialized temporal mechanisms directly into the architecture of the learning system. The slow calcium dynamics of astrocytes serve as a natural ``timer'' or ``temporal stamp'', marking recently visited states without the need for explicit history storage. This paves the way for energy-efficient hardware implementations of autonomous agents capable of real-time adaptation without computationally expensive intrinsic motivation modules.

\section{Conclusions}\label{sec_conclusion}

We have presented a spiking neuron–astrocyte network (SNAN) that resolves the exploration–exploitation dilemma through the interaction of two local plasticity mechanisms operating on distinct timescales. Spike-timing-dependent plasticity (STDP) provides long-term memory for exploiting traversed routes, while astrocytic calcium dynamics create a transient ``recently visited'' trace that promotes exploration by temporarily suppressing familiar actions. This dual-mechanism architecture enables efficient navigation in partially observable grid environments in which the agent observes only its current position, without external reward signals or global statistics. The astrocytic modulation increased goal completion from 45.9\% to 99.9\% in the open 10×10 grid and from 13.0–45.9\% to at least 96.2\% in static mazes, and shortened median routes 5.8–15-fold on task instances solved with and without astrocytes. The system proved to be robust across obstacle densities and in multi-goal tasks requiring a return to the start. Compared with classical navigation strategies, SNAN produced the shortest routes among reliable strategies in single-goal navigation in open and sparsely obstructed environments while storing 2.8--4.6 times fewer variables than Tr\'emaux's and Tarry's algorithms and Q-learning. Unlike the classical maze-solving algorithms, SNAN improved its routes through repeated experience, and after 10 learning cycles it produced the shortest routes of all reliable strategies at all tested wall fractions, although Q-learning converged to shorter routes with longer training. Owing to the self-erasing nature of astrocytic memory, SNAN became more efficient rather than less in changing mazes. With walls relocated at every step, it produced the shortest median routes of all methods at 20\% and 40\% wall density, with the largest advantage in sparse mazes. We interpret this mechanism as a form of working memory, which we term Topological-Context Memory, in which current sensory data are processed together with the context encoded locally in topologically specific neuron–astrocyte interactions.

Furthermore, we demonstrated that the STDP component can be replaced by a memristive VTEAM model calibrated to experimental device data, and that the excitatory synaptic array can be physically implemented on a memristive crossbar. The neuromorphic hardware realization achieved speed improvement and significant reduction in energy per decision compared to a conventional CPU, while maintaining equivalent task performance. These results establish astrocytic modulation as a biologically inspired, hardware-realizable mechanism for autonomous exploration in neuromorphic agents, with direct applications in energy-constrained robotics, wearable devices, and edge-AI systems.

\section*{Code availability}
The code is available at \url{https://github.com/altergot/Neuromorphic_Agent_Navigation_Control_System_with_Short-Term_Memory}.

% CRediT authorship contribution statement (taken from \credit{} of each author)
\printcredits

\section*{Declaration of competing interest}
The authors declare that they have no known competing financial interests or personal relationships that could have appeared to influence the work reported in this paper.

\section*{Funding}
This study was conducted within the framework of the scientific program of the National Center for Physics and Mathematics, section No.~9 ``Artificial intelligence and big data in technical, industrial, natural and social systems''. The memristive hardware was designed and fabricated at the facilities of the Laboratory of Memristor Nanoelectronics (state assignment for the creation of new laboratories for electronics industry No.~FSWR-2025-0006) and the Educational Electronics Design Center of Lobachevsky University (federal project ``Training of personnel and scientific foundation for electronics industry'').

\bibliographystyle{cas-model2-names}
\bibliography{refs}

@incollection{Akpan2020,
  title={Classical and Operant Conditioning---{Ivan Pavlov}; {Burrhus Skinner}},
  ISBN={9783030436209}, 
  ISSN={2366-7680}, 
  url={http://dx.doi.org/10.1007/978-3-030-43620-9\_6},
  DOI={10.1007/978-3-030-43620-9_6},
  booktitle={Science Education in Theory and Practice},
  publisher={Springer International Publishing}, 
  author={Akpan, Ben}, 
  year={2020}, 
  pages={71--84} 
}

@article{Staddon2003,
title={Operant Conditioning},
volume={54}, 
ISSN={1545-2085}, 
url={http://dx.doi.org/10.1146/annurev.psych.54.101601.145124},
DOI={10.1146/annurev.psych.54.101601.145124}, 
number={1}, 
journal={Annual Review of Psychology}, 
publisher={Annual Reviews}, 
author={Staddon, J. E. R. and Cerutti, D. T.}, 
year={2003}, 
month=feb, 
pages={115–144} 
}

@article{Barto1981,
title={Associative search network: A reinforcement learning associative memory},
volume={40},
ISSN={1432-0770},
url={http://dx.doi.org/10.1007/BF00453370}, 
DOI={10.1007/bf00453370},
number={3}, 
journal={Biological Cybernetics}, 
publisher={Springer Science and Business Media LLC}, 
author={Barto, Andrew G. and Sutton, Richard S. and Brouwer, Peter S.},
year={1981},
month=may, 
pages={201–211} }

@article{Singh2000,
title={Convergence Results for Single-Step On-Policy Reinforcement-Learning Algorithms}, 
volume={38}, 
ISSN={1573-0565}, 
url={http://dx.doi.org/10.1023/A:1007678930559},
DOI={10.1023/a:1007678930559},
number={3}, 
journal={Machine Learning},
publisher={Springer Science and Business Media LLC}, 
author={Singh, Satinder and Jaakkola, Tommi and Littman, Michael L. and Szepesvári, Csaba},
year={2000}, 
month=mar,
pages={287–308}
}

@article{Tihomirov2025,
title={Combination of reward-modulated spike-timing dependent plasticity and temporal difference long-term potentiation in actor–critic spiking neural network},
volume={90},
ISSN={1389-0417},
url={http://dx.doi.org/10.1016/j.cogsys.2025.101334},
DOI={10.1016/j.cogsys.2025.101334}, 
journal={Cognitive Systems Research}, 
publisher={Elsevier BV}, 
author={Tihomirov, Yunes and Rybka, Roman and Serenko, Alexey and Sboev, Alexander}, 
year={2025},
month=apr, 
pages={101334}
}

@article{Oh2015,
title={Action-conditional video prediction using deep networks in {Atari} games},
volume={28},
journal={Advances in Neural Information Processing Systems 28 (NIPS 2015)}, 
author={Oh, Junhyuk and Guo, Xiaoxiao and Lee, Honglak and Lewis, Richard L. and Singh, Satinder}, 
year={2015}, 
pages={2863-2871}
}

@article{Vinyals2019,
title={Grandmaster level in {StarCraft II} using multi-agent reinforcement learning},
volume={575}, 
ISSN={1476-4687}, 
url={http://dx.doi.org/10.1038/s41586-019-1724-z},
DOI={10.1038/s41586-019-1724-z}, 
number={7782}, 
journal={Nature}, 
publisher={Springer Science and Business Media LLC}, 
author={Vinyals, Oriol and Babuschkin, Igor and Czarnecki, Wojciech M. and Mathieu, Michaël and Dudzik, Andrew and Chung, Junyoung and Choi, David H. and Powell, Richard and Ewalds, Timo and Georgiev, Petko and Oh, Junhyuk and Horgan, Dan and Kroiss, Manuel and Danihelka, Ivo and Huang, Aja and Sifre, Laurent and Cai, Trevor and Agapiou, John P. and Jaderberg, Max and Vezhnevets, Alexander S. and Leblond, Rémi and Pohlen, Tobias and Dalibard, Valentin and Budden, David and Sulsky, Yury and Molloy, James and Paine, Tom L. and Gulcehre, Caglar and Wang, Ziyu and Pfaff, Tobias and Wu, Yuhuai and Ring, Roman and Yogatama, Dani and Wünsch, Dario and McKinney, Katrina and Smith, Oliver and Schaul, Tom and Lillicrap, Timothy and Kavukcuoglu, Koray and Hassabis, Demis and Apps, Chris and Silver, David}, 
year={2019}, 
month=oct, 
pages={350–354} 
}

@article{Vlasov2023,
title={Memristor-based spiking neural network with online reinforcement learning}, 
volume={166},
ISSN={0893-6080},
url={http://dx.doi.org/10.1016/j.neunet.2023.07.031}, 
DOI={10.1016/j.neunet.2023.07.031}, 
journal={Neural Networks}, 
publisher={Elsevier BV},
author={Vlasov, Danila and Minnekhanov, Anton and Rybka, Roman and Davydov, Yury and Sboev, Alexander and Serenko, Alexey and Ilyasov, Alexander and Demin, Vyacheslav},
year={2023}, 
month=sep, 
pages={512–523} 
}

@article{lee2023,
  title={{RLAIF}: Scaling reinforcement learning from human feedback with {AI} feedback},
  journal={arXiv preprint arXiv:2309.00267},
  author={Lee, Harrison and Phatale, Samrat and Mansoor, Hassan and Lu, Kellie Ren and Mesnard, Thomas and Ferret, Johan and Bishop, Colton and Hall, Ethan and Carbune, Victor and Rastogi, Abhinav},
  year={2023}
}

@book{Slivkins2019,
  title={Introduction to Multi-Armed Bandits},
  author={Slivkins, Aleksandrs},
  isbn={9781680836219},
  series={Foundations and Trends in Machine Learning Series},
  url={https://books.google.ru/books?id=6ViCzQEACAAJ},
  year={2019},
  publisher={Now Publishers}
}

@article{Auer2002,
title={Finite-time Analysis of the Multiarmed Bandit Problem},
volume={47},
ISSN={1573-0565}, 
url={http://dx.doi.org/10.1023/A:1013689704352},
DOI={10.1023/a:1013689704352},
number={2–3}, 
journal={Machine Learning},
publisher={Springer Science and Business Media LLC},
author={Auer, Peter and Cesa-Bianchi, Nicolò and Fischer, Paul},
year={2002}, 
month=may,
pages={235–256}
}

@article{Murphy2025,
title={Surveying the Space of Descriptions of a Composite System with Machine Learning},
volume={134},
ISSN={1079-7114},
url={http://dx.doi.org/10.1103/gxrh-2xsv}, 
DOI={10.1103/gxrh-2xsv}, 
number={25}, 
journal={Physical Review Letters},
publisher={American Physical Society (APS)},
author={Murphy, Kieran A. and Zhang, Yujing and Bassett, Dani S.},
year={2025}, 
month=jun,
pages={257401}
}

@article{Greff2017,
title={{LSTM}: A Search Space Odyssey},
volume={28}, 
ISSN={2162-2388},
url={http://dx.doi.org/10.1109/TNNLS.2016.2582924}, 
DOI={10.1109/tnnls.2016.2582924}, 
number={10}, 
journal={IEEE Transactions on Neural Networks and Learning Systems}, 
publisher={Institute of Electrical and Electronics Engineers (IEEE)},
author={Greff, Klaus and Srivastava, Rupesh K. and Koutnik, Jan and Steunebrink, Bas R. and Schmidhuber, Jurgen},
year={2017}, 
month=oct,
pages={2222–2232}
}

@article{Eichenbaum2000,
title={A cortical–hippocampal system for declarative memory},
volume={1}, 
ISSN={1471-0048},
url={http://dx.doi.org/10.1038/35036213}, 
DOI={10.1038/35036213}, 
number={1}, 
journal={Nature Reviews Neuroscience},
publisher={Springer Science and Business Media LLC},
author={Eichenbaum, Howard},
year={2000}, 
month=oct, 
pages={41–50} 
}

@article{Kazantsev2005, 
title={Spiking dynamics of interacting oscillatory neurons},
volume={15},
ISSN={1089-7682}, 
url={http://dx.doi.org/10.1063/1.1883866}, 
DOI={10.1063/1.1883866}, 
number={2}, 
journal={Chaos: An Interdisciplinary Journal of Nonlinear Science}, 
publisher={AIP Publishing}, 
author={Kazantsev, V. B. and Nekorkin, V. I. and Binczak, S. and Jacquir, S. and Bilbault, J. M.}, 
year={2005},
month=apr, 
pages={023103}
}

@article{Gordleeva2021,
title={Modeling Working Memory in a Spiking Neuron Network Accompanied by Astrocytes},
volume={15}, 
ISSN={1662-5102}, 
url={http://dx.doi.org/10.3389/fncel.2021.631485},
DOI={10.3389/fncel.2021.631485}, 
journal={Frontiers in Cellular Neuroscience}, 
publisher={Frontiers Media SA}, 
author={Gordleeva, Susanna Yu. and Tsybina, Yuliya A. and Krivonosov, Mikhail I. and Ivanchenko, Mikhail V. and Zaikin, Alexey A. and Kazantsev, Victor B. and Gorban, Alexander N.},
year={2021},
month=mar,
pages={631485}
}

@article{Gordleeva2025,
title={Situation-Based Neuromorphic Memory in Spiking Neuron-Astrocyte Network},
volume={36},
ISSN={2162-2388},
url={http://dx.doi.org/10.1109/TNNLS.2023.3335450}, 
DOI={10.1109/tnnls.2023.3335450}, 
number={1},
journal={IEEE Transactions on Neural Networks and Learning Systems}, 
publisher={Institute of Electrical and Electronics Engineers (IEEE)}, 
author={Gordleeva, Susanna and Tsybina, Yuliya A. and Krivonosov, Mikhail I. and Tyukin, Ivan Y. and Kazantsev, Victor B. and Zaikin, Alexey and Gorban, Alexander N.},
year={2025},
month=jan, 
pages={881–895}
}

@article{Chua1971,
title={Memristor---{T}he missing circuit element},
volume={18}, 
ISSN={0018-9324},
url={http://dx.doi.org/10.1109/TCT.1971.1083337},
DOI={10.1109/tct.1971.1083337},
number={5},
journal={IEEE Transactions on Circuit Theory},
publisher={Institute of Electrical and Electronics Engineers (IEEE)}, 
author={Chua, L.},
year={1971},
pages={507–519}
}

@article{Strukov2008,
title={The missing memristor found}, 
volume={453},
ISSN={1476-4687},
url={http://dx.doi.org/10.1038/nature06932},
DOI={10.1038/nature06932}, 
number={7191}, 
journal={Nature},
publisher={Springer Science and Business Media LLC}, 
author={Strukov, Dmitri B. and Snider, Gregory S. and Stewart, Duncan R. and Williams, R. Stanley},
year={2008},
month=may,
pages={80–83} 
}

@article{Kulagin2025,
title={Reinforcement Learning of Spiking Neural Networks Using Trace Variables for Synaptic Weights with Memristive Plasticity}, 
volume={54}, 
ISSN={1608-3415},
url={http://dx.doi.org/10.1134/S1063739725600475}, 
DOI={10.1134/s1063739725600475}, 
number={3}, 
journal={Russian Microelectronics},
publisher={Pleiades Publishing Ltd}, 
author={Kulagin, V. A. and Matsukatova, A. N. and Ryl’kov, V. V. and Demin, V. A.},
year={2025}, 
month=jun, 
pages={230–239} 
}

@article{Mikhaylov2020, 
title={Neurohybrid Memristive {CMOS}-Integrated Systems for Biosensors and Neuroprosthetics},
volume={14},
ISSN={1662-453X},
url={http://dx.doi.org/10.3389/fnins.2020.00358}, 
DOI={10.3389/fnins.2020.00358}, 
journal={Frontiers in Neuroscience},
publisher={Frontiers Media SA}, 
author={Mikhaylov, Alexey and Pimashkin, Alexey and Pigareva, Yana and Gerasimova, Svetlana and Gryaznov, Evgeny and Shchanikov, Sergey and Zuev, Anton and Talanov, Max and Lavrov, Igor and Demin, Vyacheslav and Erokhin, Victor and Lobov, Sergey and Mukhina, Irina and Kazantsev, Victor and Wu, Huaqiang and Spagnolo, Bernardo}, 
year={2020},
month=apr, 
pages={358} 
}

@article{Benjamin2014,
title={Neurogrid: A Mixed-Analog-Digital Multichip System for Large-Scale Neural Simulations},
volume={102},
ISSN={1558-2256},
url={http://dx.doi.org/10.1109/JPROC.2014.2313565}, 
DOI={10.1109/jproc.2014.2313565}, 
number={5}, 
journal={Proceedings of the IEEE},
publisher={Institute of Electrical and Electronics Engineers (IEEE)},
author={Benjamin, Ben Varkey and Gao, Peiran and McQuinn, Emmett and Choudhary, Swadesh and Chandrasekaran, Anand R. and Bussat, Jean-Marie and Alvarez-Icaza, Rodrigo and Arthur, John V. and Merolla, Paul A. and Boahen, Kwabena},
year={2014},
month=may, 
pages={699–716}
}

@article{Thrun1995,
author = {Sebastian Thrun and Tom Mitchell},
title = {Lifelong Robot Learning},
journal = {Robotics and Autonomous Systems},
year = {1995},
month = {July},
volume = {15},
number = {1},
pages = {25--46},
}

@article{Rybka2024,
title={Comparison of Bagging and Sparcity Methods for Connectivity Reduction in Spiking Neural Networks with Memristive Plasticity},
volume={8},
ISSN={2504-2289},
url={http://dx.doi.org/10.3390/bdcc8030022},
DOI={10.3390/bdcc8030022}, 
number={3}, 
journal={Big Data and Cognitive Computing}, 
publisher={MDPI AG}, 
author={Rybka, Roman and Davydov, Yury and Vlasov, Danila and Serenko, Alexey and Sboev, Alexander and Ilyin, Vyacheslav}, 
year={2024}, 
month=feb, 
pages={22}
}

@article{Vlasov2024, 
title={Spiking Neural Network Actor–Critic Reinforcement Learning with Temporal Coding and Reward-Modulated Plasticity}, 
volume={79}, 
ISSN={1934-8460},
url={http://dx.doi.org/10.3103/S0027134924702400}, 
DOI={10.3103/s0027134924702400},
number={S2}, 
journal={Moscow University Physics Bulletin},
publisher={Allerton Press}, 
author={Vlasov, D. S. and Rybka, R. B. and Serenko, A. V. and Sboev, A. G.}, 
year={2024}, 
month=dec, 
pages={S944–S952}
}

@article{ULLAH2006,
  doi = {10.1016/j.ceca.2005.10.009},
  url = {https://doi.org/10.1016/j.ceca.2005.10.009},
  year = {2006},
  month = mar,
  publisher = {Elsevier {BV}},
  volume = {39},
  number = {3},
  pages = {197--208},
  author = {G Ullah and P Jung and A Cornell-Bell},
  title = {Anti-phase calcium oscillations in astrocytes via inositol (1,  4,  5)-trisphosphate regeneration},
  journal = {Cell Calcium}
}

@article{Kvatinsky2015,
  title = {{VTEAM}: A General Model for Voltage-Controlled Memristors},
  volume = {62},
  ISSN = {1558-3791},
  url = {http://dx.doi.org/10.1109/TCSII.2015.2433536},
  DOI = {10.1109/tcsii.2015.2433536},
  number = {8},
  journal = {IEEE Transactions on Circuits and Systems II: Express Briefs},
  publisher = {Institute of Electrical and Electronics Engineers (IEEE)},
  author = {Kvatinsky,  Shahar and Ramadan,  Misbah and Friedman,  Eby G. and Kolodny,  Avinoam},
  year = {2015},
  month = aug,
  pages = {786–790}
}

@article{Zenke2018,
title={{SuperSpike}: Supervised Learning in Multilayer Spiking Neural Networks}, 
volume={30}, 
ISSN={1530-888X},
url={http://dx.doi.org/10.1162/neco\_a\_01086}, 
DOI={10.1162/neco_a_01086},
number={6}, 
journal={Neural Computation},
publisher={MIT Press - Journals},
author={Zenke, Friedemann and Ganguli, Surya},
year={2018}, 
month=jun,
pages={1514–1541} 
}

@article{Banino2018,
title={Vector-based navigation using grid-like representations in artificial agents}, 
volume={557}, 
ISSN={1476-4687}, 
url={http://dx.doi.org/10.1038/s41586-018-0102-6}, 
DOI={10.1038/s41586-018-0102-6}, 
number={7705}, 
journal={Nature},
publisher={Springer Science and Business Media LLC}, 
author={Banino, Andrea and Barry, Caswell and Uria, Benigno and Blundell, Charles and Lillicrap, Timothy and Mirowski, Piotr and Pritzel, Alexander and Chadwick, Martin J. and Degris, Thomas and Modayil, Joseph and Wayne, Greg and Soyer, Hubert and Viola, Fabio and Zhang, Brian and Goroshin, Ross and Rabinowitz, Neil and Pascanu, Razvan and Beattie, Charlie and Petersen, Stig and Sadik, Amir and Gaffney, Stephen and King, Helen and Kavukcuoglu, Koray and Hassabis, Demis and Hadsell, Raia and Kumaran, Dharshan}, 
year={2018}, 
month=may, 
pages={429–433} 
}

@inproceedings{Bellemare2016,
 author = {Bellemare, Marc and Srinivasan, Sriram and Ostrovski, Georg and Schaul, Tom and Saxton, David and Munos, Remi},
 booktitle = {Advances in Neural Information Processing Systems},
 editor = {D. Lee and M. Sugiyama and U. Luxburg and I. Guyon and R. Garnett},
 publisher = {Curran Associates},
 title = {Unifying Count-Based Exploration and Intrinsic Motivation},
 volume = {29},
 pages = {1471--1479},
 year = {2016}
}

@article{Fremaux2016,
title={Neuromodulated Spike-Timing-Dependent Plasticity, and Theory of Three-Factor Learning Rules}, 
volume={9}, 
ISSN={1662-5110},
url={http://dx.doi.org/10.3389/fncir.2015.00085}, 
DOI={10.3389/fncir.2015.00085}, 
journal={Frontiers in Neural Circuits}, 
publisher={Frontiers Media SA}, 
author={Frémaux, Nicolas and Gerstner, Wulfram}, 
year={2016}, 
month=jan 
}

@article{Liu2019, 
title={Exploring Self-Repair in a Coupled Spiking Astrocyte Neural Network},
volume={30},
ISSN={2162-2388},
url={http://dx.doi.org/10.1109/TNNLS.2018.2854291},
DOI={10.1109/tnnls.2018.2854291}, 
number={3}, 
journal={IEEE Transactions on Neural Networks and Learning Systems}, 
publisher={Institute of Electrical and Electronics Engineers (IEEE)}, 
author={Liu, Junxiu and Mcdaid, Liam J. and Harkin, Jim and Karim, Shvan and Johnson, Anju P. and Millard, Alan G. and Hilder, James and Halliday, David M. and Tyrrell, Andy M. and Timmis, Jon}, 
year={2019}, 
month=mar,
pages={865–875}
}

@article{Sutton1999,
  title={Between {MDPs} and semi-{MDPs}: A framework for temporal abstraction in reinforcement learning},
  author={Sutton, Richard S and Precup, Doina and Singh, Satinder},
  journal={Artificial Intelligence},
  volume={112},
  number={1-2},
  pages={181--211},
  year={1999},
  publisher={Elsevier}
}

@article{Asch2025,
title={Basal ganglia deep brain stimulation restores cognitive flexibility and exploration-exploitation balance disrupted by {NMDA-R} antagonism},
volume={16},
ISSN={2041-1723},
url={http://dx.doi.org/10.1038/s41467-025-60044-5},
DOI={10.1038/s41467-025-60044-5},
number={1}, 
journal={Nature Communications},
publisher={Springer Science and Business Media LLC}, 
author={Asch, Nir and Rahamim, Noa and Morozov, Anna and Werner-Reiss, Uri and Israel, Zvi and Paz, Rony and Bergman, Hagai}, 
year={2025}, 
month=may
}

@article{Bacon2017,
title={The Option-Critic Architecture},
volume={31},
ISSN={2159-5399},
url={http://dx.doi.org/10.1609/aaai.v31i1.10916},
DOI={10.1609/aaai.v31i1.10916},
number={1}, 
journal={Proceedings of the AAAI Conference on Artificial Intelligence},
publisher={Association for the Advancement of Artificial Intelligence (AAAI)},
author={Bacon, Pierre-Luc and Harb, Jean and Precup, Doina}, 
year={2017}, 
month=feb 
}

@article{Burda2018,
  title={Exploration by random network distillation},
  author={Burda, Yuri and Edwards, Harrison and Storkey, Amos and Klimov, Oleg},
  journal={arXiv preprint arXiv:1810.12894},
  year={2018}
}

@article{Duan2016,
  title={{RL}$^2$: Fast reinforcement learning via slow reinforcement learning},
  author={Duan, Yan and Schulman, John and Chen, Xi and Bartlett, Peter L and Sutskever, Ilya and Abbeel, Pieter},
  journal={arXiv preprint arXiv:1611.02779},
  year={2016}
}

@article{Dzhivelikian2022,
title={Hierarchical intrinsically motivated agent planning behavior with dreaming in grid environments},
volume={9},
ISSN={2198-4026}, 
url={http://dx.doi.org/10.1186/s40708-022-00156-6}, 
DOI={10.1186/s40708-022-00156-6}, 
number={1}, 
journal={Brain Informatics},
publisher={Springer Science and Business Media LLC},
author={Dzhivelikian, Evgenii and Latyshev, Artem and Kuderov, Petr and Panov, Aleksandr I.}, 
year={2022}, 
month=apr 
}

@article{Lai1985,
title={Asymptotically efficient adaptive allocation rules},
volume={6}, 
ISSN={0196-8858},
url={http://dx.doi.org/10.1016/0196-8858(85)90002-8}, 
DOI={10.1016/0196-8858(85)90002-8},
number={1}, 
journal={Advances in Applied Mathematics}, 
publisher={Elsevier BV}, 
author={Lai, T.L and Robbins, Herbert}, 
year={1985}, 
month=mar, 
pages={4–22} 
}

@book{Luce1959,
  title={Individual Choice Behavior: A Theoretical Analysis},
  author={Luce, R.D.},
  lccn={59009346},
  url={https://books.google.ru/books?id=a80DAQAAIAAJ},
  year={1959},
  publisher={Wiley}
}

@inproceedings{Norman2024,
series={NeurIPS 2024},
title={{First-Explore}, then Exploit: Meta-Learning to Solve Hard Exploration-Exploitation Trade-Offs},
url={http://dx.doi.org/10.52202/079017-0864},
DOI={10.52202/079017-0864},
booktitle={Advances in Neural Information Processing Systems 37},
publisher={Neural Information Processing Systems Foundation, Inc. (NeurIPS)},
author = {B Norman and J Clune},
year={2024}, 
pages={27490–27528},
collection={NeurIPS 2024}
}

@InProceedings{Ostrovski2017,
  title = 	 {Count-Based Exploration with Neural Density Models},
  author =       {Georg Ostrovski and Marc G. Bellemare and A{\"a}ron van den Oord and R{\'e}mi Munos},
  booktitle = 	 {Proceedings of the 34th International Conference on Machine Learning},
  pages = 	 {2721--2730},
  year = 	 {2017},
  editor = 	 {Precup, Doina and Teh, Yee Whye},
  volume = 	 {70},
  series = 	 {Proceedings of Machine Learning Research},
  month = 	 {06--11 Aug},
  publisher =    {PMLR},
  url = 	 {https://proceedings.mlr.press/v70/ostrovski17a.html},
}

@inproceedings{Pathak2017,
  title={Curiosity-driven exploration by self-supervised prediction},
  author={Pathak, Deepak and Agrawal, Pulkit and Efros, Alexei A and Darrell, Trevor},
  booktitle={International Conference on Machine Learning},
  pages={2778--2787},
  year={2017},
  organization={PMLR}
}

@article{Pechac2024,
title={Self-supervised network distillation: An effective approach to exploration in sparse reward environments}, 
volume={599}, 
ISSN={0925-2312},
url={http://dx.doi.org/10.1016/j.neucom.2024.128033},
DOI={10.1016/j.neucom.2024.128033}, 
journal={Neurocomputing},
publisher={Elsevier BV},
author={Pecháč, Matej and Chovanec, Michal and Farkaš, Igor},
year={2024},
month=sep,
pages={128033} 
}

@book{Sutton1998,
  title={Reinforcement learning: An introduction},
  author={Sutton, Richard S and Barto, Andrew G},
  year={1998},
  publisher={MIT Press},
  address={Cambridge, MA}
}

@article{Thompson1933, 
title={On the Likelihood that One Unknown Probability Exceeds Another in View of the Evidence of Two Samples},
volume={25},
ISSN={0006-3444}, 
url={http://dx.doi.org/10.2307/2332286}, 
DOI={10.2307/2332286}, 
number={3/4}, 
journal={Biometrika},
publisher={JSTOR}, 
author={Thompson, William R.},
year={1933}, 
month=dec,
pages={285}
}

@phdthesis{Watkins1989,
  title={Learning from delayed rewards},
  author={Watkins, Christopher John Cornish Hellaby},
  year={1989},
  school={King's College, University of Cambridge}
}

@article{Zintgraf2020,
  title={{VariBAD}: Variational {Bayes}-Adaptive Deep {RL} via Meta-Learning},
  author={Zintgraf, Luisa and Schulze, Sebastian and Lu, Cong and Feng, Leo and Igl, Maximilian and Shiarlis, Kyriacos and Gal, Yarin and Hofmann, Katja and Whiteson, Shimon},
  journal={Journal of Machine Learning Research},
  volume={22},
  number={289},
  pages={1--39},
  year={2021}
}

@inproceedings{shchanikov2023modeling,
  title={Modeling and hardware implementation of vector-matrix multiplier based on 32x8 {1T1R} memristive crossbar array},
  author={Shchanikov, Sergey and Korolev, Leonid and Bordanov, Ilya and Belov, Alexey and Gryaznov, Evgeny and Mikhaylov, Alexey},
  booktitle={2023 7th Scientific School Dynamics of Complex Networks and their Applications (DCNA)},
  pages={249--251},
  year={2023},
  organization={IEEE}
}

@article{mikhaylov2020multilayer,
  title={Multilayer metal-oxide memristive device with stabilized resistive switching},
  author={Mikhaylov, Alexey and Belov, Alexey and Korolev, Dmitry and Antonov, Ivan and Kotomina, Valentina and Kotina, Alina and Gryaznov, Evgeny and Sharapov, Alexander and Koryazhkina, Maria and Kryukov, Ruslan and others},
  journal={Advanced Materials Technologies},
  volume={5},
  number={1},
  pages={1900607},
  year={2020},
  publisher={Wiley Online Library}
}

@article{mikhaylov2023neuromorphic,
  title={Neuromorphic computing based on {CMOS}-integrated memristive arrays: current state and perspectives},
  author={Mikhaylov, Alexey N and Gryaznov, Evgeny G and Koryazhkina, Maria N and Bordanov, Ilya A and Shchanikov, Sergey A and Telminov, Oleg A and Kazantsev, Victor B},
  journal={Supercomputing Frontiers and Innovations},
  volume={10},
  number={2},
  pages={77--103},
  year={2023}
}

@article{liu2025memristor,
  title={A memristor-based adaptive neuromorphic decoder for brain--computer interfaces},
  author={Liu, Zhengwu and Mei, Jie and Tang, Jianshi and Xu, Minpeng and Gao, Bin and Wang, Kun and Ding, Sanchuang and Liu, Qi and Qin, Qi and Chen, Weize and others},
  journal={Nature Electronics},
  volume={8},
  number={4},
  pages={362--372},
  year={2025},
  publisher={Nature Publishing Group UK London}
}

@misc{intelbenchmark,
  author = {{PassMark Software}},
  year = {2026},
  title = {{Intel Core i5-12450H} Benchmark},
  howpublished = {\url{https://www.cpubenchmark.net/cpu.php?cpu=Intel+Core+i5-12450H\&id=4727}},
  note = {Accessed: 2026-03-04}
}

@misc{shchanikov_memriboard,
    author = {{Neurocomputer}},
    year = {2026},
    title = {{MemriBoard} Framework},
    howpublished = {\url{https://github.com/neurocomputer/MemriBoard}},
    note = {Accessed: 2026-03-04}
}

@article{Durstewitz2025,
title={What neuroscience can tell {AI} about learning in continuously changing environments}, 
volume={7}, 
ISSN={2522-5839}, 
url={http://dx.doi.org/10.1038/s42256-025-01146-z}, 
DOI={10.1038/s42256-025-01146-z}, 
number={12},
journal={Nature Machine Intelligence},
publisher={Springer Science and Business Media LLC}, 
author={Durstewitz, Daniel and Averbeck, Bruno and Koppe, Georgia}, 
year={2025}, 
month=nov, 
pages={1897–1912}
}

@article{Chowdhury2025,
title={Neuromorphic computing for robotic vision: algorithms to hardware advances},
volume={4},
ISSN={2731-3395}, 
url={http://dx.doi.org/10.1038/s44172-025-00492-5}, 
DOI={10.1038/s44172-025-00492-5},
number={1}, 
journal={Communications Engineering}, 
publisher={Springer Science and Business Media LLC}, 
author={Chowdhury, Sayeed Shafayet and Sharma, Deepika and Kosta, Adarsh and Roy, Kaushik},
year={2025},
month=aug 
}

@article{Novo2024,
title={Neuromorphic Perception and Navigation for Mobile Robots: A Review},
volume={56},
ISSN={1557-7341},
url={http://dx.doi.org/10.1145/3656469},
DOI={10.1145/3656469},
number={10}, 
journal={ACM Computing Surveys}, 
publisher={Association for Computing Machinery (ACM)}, 
author={Novo, Alvaro and Lobon, Francisco and Garcia de Marina, Hector and Romero, Samuel and Barranco, Francisco},
year={2024}, 
month=may,
pages={1–37} 
}

@article{Maass1997, 
title={Networks of spiking neurons: The third generation of neural network models}, 
volume={10},
ISSN={0893-6080},
url={http://dx.doi.org/10.1016/S0893-6080(97)00011-7}, 
DOI={10.1016/s0893-6080(97)00011-7},
number={9}, 
journal={Neural Networks},
publisher={Elsevier BV},
author={Maass, Wolfgang},
year={1997}, 
month=dec, 
pages={1659–1671}
}

@article{Roy2019,
title={Towards spike-based machine intelligence with neuromorphic computing},
volume={575},
ISSN={1476-4687},
url={http://dx.doi.org/10.1038/s41586-019-1677-2}, 
DOI={10.1038/s41586-019-1677-2}, 
number={7784}, 
journal={Nature}, 
publisher={Springer Science and Business Media LLC}, 
author={Roy, Kaushik and Jaiswal, Akhilesh and Panda, Priyadarshini},
year={2019}, 
month=nov, 
pages={607–617} 
}

@article{Davies2018, 
title={Loihi: A Neuromorphic Manycore Processor with On-Chip Learning}, 
volume={38}, 
ISSN={1937-4143},
url={http://dx.doi.org/10.1109/MM.2018.112130359},
DOI={10.1109/mm.2018.112130359},
number={1}, 
journal={IEEE Micro},
publisher={Institute of Electrical and Electronics Engineers (IEEE)}, 
author={Davies, Mike and Srinivasa, Narayan and Lin, Tsung-Han and Chinya, Gautham and Cao, Yongqiang and Choday, Sri Harsha and Dimou, Georgios and Joshi, Prasad and Imam, Nabil and Jain, Shweta and Liao, Yuyun and Lin, Chit-Kwan and Lines, Andrew and Liu, Ruokun and Mathaikutty, Deepak and McCoy, Steven and Paul, Arnab and Tse, Jonathan and Venkataramanan, Guruguhanathan and Weng, Yi-Hsin and Wild, Andreas and Yang, Yoonseok and Wang, Hong},
year={2018}, 
month=jan, 
pages={82–99} 
}

@article{Akopyan2015,
title={{TrueNorth}: Design and Tool Flow of a 65~{mW} 1 Million Neuron Programmable Neurosynaptic Chip}, 
volume={34}, 
ISSN={1937-4151}, 
url={http://dx.doi.org/10.1109/TCAD.2015.2474396}, 
DOI={10.1109/tcad.2015.2474396}, 
number={10}, 
journal={IEEE Transactions on Computer-Aided Design of Integrated Circuits and Systems}, 
publisher={Institute of Electrical and Electronics Engineers (IEEE)},
author={Akopyan, Filipp and Sawada, Jun and Cassidy, Andrew and Alvarez-Icaza, Rodrigo and Arthur, John and Merolla, Paul and Imam, Nabil and Nakamura, Yutaka and Datta, Pallab and Nam, Gi-Joon and Taba, Brian and Beakes, Michael and Brezzo, Bernard and Kuang, Jente B. and Manohar, Rajit and Risk, William P. and Jackson, Bryan and Modha, Dharmendra S.},
year={2015},
month=oct, 
pages={1537–1557} 
}

@article{Furber2014,
title={The {SpiNNaker} Project}, 
volume={102}, 
ISSN={1558-2256}, 
url={http://dx.doi.org/10.1109/JPROC.2014.2304638},
DOI={10.1109/jproc.2014.2304638}, 
number={5}, 
journal={Proceedings of the IEEE}, 
publisher={Institute of Electrical and Electronics Engineers (IEEE)},
author={Furber, Steve B. and Galluppi, Francesco and Temple, Steve and Plana, Luis A.},
year={2014},
month=may, 
pages={652–665}
}

@article{Benosman2014,
title={Event-Based Visual Flow},
volume={25}, 
ISSN={2162-2388},
url={http://dx.doi.org/10.1109/TNNLS.2013.2273537}, 
DOI={10.1109/tnnls.2013.2273537}, 
number={2}, 
journal={IEEE Transactions on Neural Networks and Learning Systems},
publisher={Institute of Electrical and Electronics Engineers (IEEE)},
author={Benosman, Ryad and Clercq, Charles and Lagorce, Xavier and Sio-Hoi Ieng and Bartolozzi, Chiara},
year={2014},
month=feb,
pages={407–417} 
}

@incollection{Barranco2015,
title={Bio-inspired Motion Estimation with Event-Driven Sensors},
ISBN={9783319192581},
ISSN={1611-3349}, 
url={http://dx.doi.org/10.1007/978-3-319-19258-1\_27}, 
DOI={10.1007/978-3-319-19258-1_27}, 
booktitle={Advances in Computational Intelligence}, 
publisher={Springer International Publishing},
author={Barranco, Francisco and Fermuller, Cornelia and Aloimonos, Yiannis}, 
year={2015}, 
pages={309--321} 
}

@inproceedings{Rebecq2017,
series={BMVC 2017}, 
title={Real-time Visual-Inertial Odometry for Event Cameras using Keyframe-based Nonlinear Optimization},
url={http://dx.doi.org/10.5244/C.31.16}, 
DOI={10.5244/c.31.16},
booktitle={Proceedings of the British Machine Vision Conference 2017}, 
publisher={British Machine Vision Association},
author={Rebecq, Henri and Horstschaefer, Timo and Scaramuzza, Davide},
year={2017},
pages={16.1--16.12}
}

@article{Zhou2021,
title={Event-Based Stereo Visual Odometry},
volume={37}, 
ISSN={1941-0468},
url={http://dx.doi.org/10.1109/TRO.2021.3062252}, 
DOI={10.1109/tro.2021.3062252},
number={5},
journal={IEEE Transactions on Robotics},
publisher={Institute of Electrical and Electronics Engineers (IEEE)}, 
author={Zhou, Yi and Gallego, Guillermo and Shen, Shaojie},
year={2021}, 
month=oct, 
pages={1433–1450}
}

@article{Rancon2022, 
title={{StereoSpike}: Depth Learning With a Spiking Neural Network}, 
volume={10}, 
ISSN={2169-3536},
url={http://dx.doi.org/10.1109/ACCESS.2022.3226484},
DOI={10.1109/access.2022.3226484}, 
journal={IEEE Access},
publisher={Institute of Electrical and Electronics Engineers (IEEE)},
author={Rancon, Ulysse and Cuadrado-Anibarro, Javier and Cottereau, Benoit R. and Masquelier, Timothee},
year={2022},
pages={127428–127439}
}

@incollection{Lee2020, 
title={{Spike-FlowNet}: Event-Based Optical Flow Estimation with Energy-Efficient Hybrid Neural Networks},
ISBN={9783030585266},
ISSN={1611-3349},
url={http://dx.doi.org/10.1007/978-3-030-58526-6\_22},
DOI={10.1007/978-3-030-58526-6_22}, 
booktitle={Computer Vision – ECCV 2020},
publisher={Springer International Publishing},
author={Lee, Chankyu and Kosta, Adarsh Kumar and Zhu, Alex Zihao and Chaney, Kenneth and Daniilidis, Kostas and Roy, Kaushik},
year={2020}, 
pages={366--382}
}

@inproceedings{Kosta2023, 
title={{Adaptive-SpikeNet}: Event-based Optical Flow Estimation using Spiking Neural Networks with Learnable Neuronal Dynamics},
url={http://dx.doi.org/10.1109/ICRA48891.2023.10160551},
DOI={10.1109/icra48891.2023.10160551},
booktitle={2023 IEEE International Conference on Robotics and Automation (ICRA)}, 
publisher={IEEE},
author={Kosta, Adarsh Kumar and Roy, Kaushik},
year={2023}, 
month=may,
pages={6021–6027}
}

@article{Ball2013,
title={{OpenRatSLAM}: an open source brain-based {SLAM} system}, 
volume={34}, 
ISSN={1573-7527}, 
url={http://dx.doi.org/10.1007/s10514-012-9317-9},
DOI={10.1007/s10514-012-9317-9},
number={3}, 
journal={Autonomous Robots},
publisher={Springer Science and Business Media LLC}, 
author={Ball, David and Heath, Scott and Wiles, Janet and Wyeth, Gordon and Corke, Peter and Milford, Michael},
year={2013},
month=feb, 
pages={149–176}
}

@article{Milford2008, 
title={Mapping a Suburb With a Single Camera Using a Biologically Inspired {SLAM} System}, 
volume={24},
ISSN={1941-0468},
url={http://dx.doi.org/10.1109/TRO.2008.2004520}, 
DOI={10.1109/tro.2008.2004520},
number={5}, 
journal={IEEE Transactions on Robotics},
publisher={Institute of Electrical and Electronics Engineers (IEEE)},
author={Milford, M.J. and Wyeth, G.F.}, 
year={2008}, 
month=oct, 
pages={1038–1053}
}

@article{Yu2019, 
title={{NeuroSLAM}: a brain-inspired {SLAM} system for {3D} environments},
volume={113},
ISSN={1432-0770},
url={http://dx.doi.org/10.1007/s00422-019-00806-9}, 
DOI={10.1007/s00422-019-00806-9},
number={5–6}, 
journal={Biological Cybernetics}, 
publisher={Springer Science and Business Media LLC},
author={Yu, Fangwen and Shang, Jianga and Hu, Youjian and Milford, Michael}, 
year={2019},
month=sep, 
pages={515–545}
}

@inproceedings{Edvardsen2017,
title={Long-range navigation by path integration and decoding of grid cells in a neural network}, 
url={http://dx.doi.org/10.1109/IJCNN.2017.7966406}, 
DOI={10.1109/ijcnn.2017.7966406}, 
booktitle={2017 International Joint Conference on Neural Networks (IJCNN)}, 
publisher={IEEE}, 
author={Edvardsen, Vegard},
year={2017},
month=may, 
pages={4348–4355} 
}

@article{Chen2021,
title={A Positioning Method Based on Place Cells and Head-Direction Cells for Inertial/Visual Brain-Inspired Navigation System},
volume={21}, 
ISSN={1424-8220},
url={http://dx.doi.org/10.3390/s21237988},
DOI={10.3390/s21237988}, 
number={23}, 
journal={Sensors}, 
publisher={MDPI AG},
author={Chen, Yudi and Xiong, Zhi and Liu, Jianye and Yang, Chuang and Chao, Lijun and Peng, Yang}, 
year={2021}, 
month=nov,
pages={7988}
}

@article{Samsonovich1997,
title={Path Integration and Cognitive Mapping in a Continuous Attractor Neural Network Model}, 
volume={17}, 
ISSN={1529-2401},
url={http://dx.doi.org/10.1523/JNEUROSCI.17-15-05900.1997},
DOI={10.1523/jneurosci.17-15-05900.1997}, 
number={15},
journal={The Journal of Neuroscience}, 
publisher={Society for Neuroscience},
author={Samsonovich, Alexei and McNaughton, Bruce L.},
year={1997}, 
month=aug,
pages={5900–5920}
}

@article{Edwards1990,
title={On the statistics of binned neural point processes: the {Bernoulli} approximation and {AR} representation of the {PST} histogram},
volume={64},
ISSN={1432-0770},
url={http://dx.doi.org/10.1007/BF02331344}, 
DOI={10.1007/bf02331344},
number={2}, 
journal={Biological Cybernetics},
publisher={Springer Science and Business Media LLC},
author={Edwards, B. W. and Wakefield, G. H.}, 
year={1990},
month=dec,
pages={145–153} 
}

@article{Gordleeva2012,
title={Bi-directional astrocytic regulation of neuronal activity within a network}, 
volume={6},
ISSN={1662-5188},
url={http://dx.doi.org/10.3389/fncom.2012.00092},
DOI={10.3389/fncom.2012.00092}, 
journal={Frontiers in Computational Neuroscience}, 
publisher={Frontiers Media SA}, 
author={Gordleeva, S. Yu and Stasenko, S. V. and Semyanov, A. V. and Dityatev, A. E. and Kazantsev, V. B.},
year={2012} 
}

@article{Santello2019,
title={Astrocyte function from information processing to cognition and cognitive impairment}, 
volume={22}, 
ISSN={1546-1726}, 
url={http://dx.doi.org/10.1038/s41593-018-0325-8}, 
DOI={10.1038/s41593-018-0325-8}, 
number={2},
journal={Nature Neuroscience},
publisher={Springer Science and Business Media LLC},
author={Santello, Mirko and Toni, Nicolas and Volterra, Andrea}, 
year={2019}, 
month=jan,
pages={154–166} 
}

@article{Pabst2016,
title={Astrocyte Intermediaries of Septal Cholinergic Modulation in the Hippocampus},
volume={90},
ISSN={0896-6273}, 
url={http://dx.doi.org/10.1016/j.neuron.2016.04.003},
DOI={10.1016/j.neuron.2016.04.003}, 
number={4}, 
journal={Neuron},
publisher={Elsevier BV}, 
author={Pabst, Milan and Braganza, Oliver and Dannenberg, Holger and Hu, Wen and Pothmann, Leonie and Rosen, Jurij and Mody, Istvan and van Loo, Karen and Deisseroth, Karl and Becker, Albert J. and Schoch, Susanne and Beck, Heinz}, 
year={2016},
month=may,
pages={853–865}
}

@article{Semyanov2020, 
title={Making sense of astrocytic calcium signals — from acquisition to interpretation},
volume={21},
ISSN={1471-0048},
url={http://dx.doi.org/10.1038/s41583-020-0361-8},
DOI={10.1038/s41583-020-0361-8},
number={10}, 
journal={Nature Reviews Neuroscience},
publisher={Springer Science and Business Media LLC}, 
author={Semyanov, Alexey and Henneberger, Christian and Agarwal, Amit},
year={2020}, 
month=sep,
pages={551–564}
}

@inproceedings{Schneider2024,
title={A Scalable Platform for Robot Learning and Physical Skill Data Collection}, 
url={http://dx.doi.org/10.1109/IROS58592.2024.10801516},
DOI={10.1109/iros58592.2024.10801516},
booktitle={2024 IEEE/RSJ International Conference on Intelligent Robots and Systems (IROS)},
publisher={IEEE},
author={Schneider, Samuel and Wu, Yansong and Johannsmeier, Lars and Wu, Fan and Haddadin, Sami},
year={2024}, 
month=oct,
pages={5925–5932}
}

@article{Matsukatova2023, 
title={Combination of Organic-Based Reservoir Computing and Spiking Neuromorphic Systems for a Robust and Efficient Pattern Classification},
volume={5}, 
ISSN={2640-4567},
url={http://dx.doi.org/10.1002/aisy.202200407},
DOI={10.1002/aisy.202200407}, 
number={6}, 
journal={Advanced Intelligent Systems},
publisher={Wiley}, 
author={Matsukatova, Anna N. and Prudnikov, Nikita V. and Kulagin, Vsevolod A. and Battistoni, Silvia and Minnekhanov, Anton A. and Trofimov, Andrey D. and Nesmelov, Aleksandr A. and Zavyalov, Sergey A. and Malakhova, Yulia N. and Parmeggiani, Matteo and Ballesio, Alberto and Marasso, Simone Luigi and Chvalun, Sergey N. and Demin, Vyacheslav A. and Emelyanov, Andrey V. and Erokhin, Victor}, 
year={2023},
month=mar 
}

@article{Matsukatova2022, 
title={Nanocomposite parylene-{C} memristors with embedded {Ag} nanoparticles for biomedical data processing},
volume={102},
ISSN={1566-1199},
url={http://dx.doi.org/10.1016/j.orgel.2022.106455},
DOI={10.1016/j.orgel.2022.106455}, 
journal={Organic Electronics}, 
publisher={Elsevier BV}, 
author={Matsukatova, Anna N. and Emelyanov, Andrey V. and Kulagin, Vsevolod A. and Vdovichenko, Artem Yu and Minnekhanov, Anton A. and Demin, Vyacheslav A.}, year={2022},
month=mar, 
pages={106455}
}

@article{Lovasz1993,
  title={Random walks on graphs: A survey},
  author={Lov{\'a}sz, L{\'a}szl{\'o}},
  journal={Combinatorics, Paul Erd{\H{o}}s is Eighty},
  volume={2},
  number={1},
  pages={1--46},
  year={1993}
}

@article{Alon2007,
  title={Non-backtracking random walks mix faster},
  author={Alon, Noga and Benjamini, Itai and Lubetzky, Eyal and Sodin, Sasha},
  journal={Communications in Contemporary Mathematics},
  volume={9},
  number={4},
  pages={585--603},
  year={2007}
}

@article{Lumelsky1987,
  title={Path-planning strategies for a point mobile automaton moving amidst unknown obstacles of arbitrary shape},
  author={Lumelsky, Vladimir J. and Stepanov, Alexander A.},
  journal={Algorithmica},
  volume={2},
  pages={403--430},
  year={1987},
  doi={10.1007/BF01840369}
}

@book{Abelson1980,
  title={Turtle Geometry: The Computer as a Medium for Exploring Mathematics},
  author={Abelson, Harold and diSessa, Andrea A.},
  publisher={MIT Press},
  address={Cambridge, MA},
  year={1980}
}

@book{Lucas1882,
  title={R{\'e}cr{\'e}ations Math{\'e}matiques},
  author={Lucas, {\'E}douard},
  volume={1},
  publisher={Gauthier-Villars},
  address={Paris},
  year={1882}
}

@book{Even2011,
  title={Graph Algorithms},
  author={Even, Shimon},
  edition={2},
  publisher={Cambridge University Press},
  address={Cambridge},
  year={2011}
}

@article{Tarry1895,
  title={Le probl{\`e}me des labyrinthes},
  author={Tarry, Gaston},
  journal={Nouvelles Annales de Math{\'e}matiques, S{\'e}rie 3},
  volume={14},
  pages={187--190},
  year={1895}
}

@article{Watkins1992,
  title={Q-learning},
  author={Watkins, Christopher J. C. H. and Dayan, Peter},
  journal={Machine Learning},
  volume={8},
  pages={279--292},
  year={1992},
  doi={10.1007/BF00992698}
}

@article{Wilcoxon1945,
  title={Individual comparisons by ranking methods},
  author={Wilcoxon, Frank},
  journal={Biometrics Bulletin},
  volume={1},
  number={6},
  pages={80--83},
  year={1945},
  doi={10.2307/3001968}
}

@article{Klein2000,
  doi = {10.1016/S1364-6613(00)01452-2},
  url = {https://doi.org/10.1016/S1364-6613(00)01452-2},
  year = {2000},
  month = apr,
  publisher = {Elsevier {BV}},
  volume = {4},
  number = {4},
  pages = {138--147},
  author = {R M Klein},
  title = {Inhibition of return},
  journal = {Trends in Cognitive Sciences}
}

@article{Glover1989,
  doi = {10.1287/ijoc.1.3.190},
  url = {https://doi.org/10.1287/ijoc.1.3.190},
  year = {1989},
  publisher = {Institute for Operations Research and the Management Sciences ({INFORMS})},
  volume = {1},
  number = {3},
  pages = {190--206},
  author = {F Glover},
  title = {Tabu Search{\textemdash}{Part I}},
  journal = {{ORSA} Journal on Computing}
}

@article{Wagner1999,
  doi = {10.1109/70.795795},
  url = {https://doi.org/10.1109/70.795795},
  year = {1999},
  month = oct,
  publisher = {Institute of Electrical and Electronics Engineers ({IEEE})},
  volume = {15},
  number = {5},
  pages = {918--933},
  author = {I A Wagner and M Lindenbaum and A M Bruckstein},
  title = {Distributed covering by ant-robots using evaporating traces},
  journal = {{IEEE} Transactions on Robotics and Automation}
}

\end{document}